\documentclass[aps,prb,twocolumn,superscriptaddress,notitlepage,showpacs]{revtex4-2}

\usepackage{epsfig}
\usepackage{amssymb} 
\usepackage{amsfonts} 
\usepackage{mathtools} 
\usepackage{dcolumn} 
\usepackage{graphicx} 
\usepackage{color}  
\usepackage{bm}  
\usepackage{comment} 
\usepackage{units}
\usepackage{sidecap} 
\usepackage{textcomp} 

\usepackage{array} 

\begin{document}
\title{Field-angle evolution of the superconducting and magnetic phases of UTe$_2$ around the $b$ axis}

\author{Sylvia K. Lewin}
\affiliation{NIST Center for Neutron Research, National Institute of Standards and Technology, Gaithersburg, MD 20899, USA}
\affiliation{Department of Physics, Quantum Materials Center, University of Maryland, College Park, MD 20742, USA}

\author{Josephine J. Yu}
\affiliation{Department of Applied Physics, Stanford University, Stanford, California 94305, USA}

\author{Corey E. Frank}
\affiliation{NIST Center for Neutron Research, National Institute of Standards and Technology, Gaithersburg, MD 20899, USA}
\affiliation{Department of Physics, Quantum Materials Center, University of Maryland, College Park, MD 20742, USA}

\author{David Graf}
\affiliation{National High Magnetic Field Laboratory, Florida State University, Tallahassee, FL 32313, USA}

\author{Patrick Chen}
\affiliation{Department of Physics, Quantum Materials Center, University of Maryland, College Park, MD 20742, USA}

\author{Sheng Ran}
\affiliation{Department of Physics, Washington University in St. Louis, St. Louis, MO 63130, USA}

\author{Yun Suk Eo}
\affiliation{Department of Physics, Quantum Materials Center, University of Maryland, College Park, MD 20742, USA}

\author{Johnpierre Paglione}
\affiliation{Department of Physics, Quantum Materials Center, University of Maryland, College Park, MD 20742, USA}
\affiliation{Canadian Institute for Advanced Research, Toronto, Ontario M5G 1Z8, Canada}

\author{S. Raghu}
\affiliation{Stanford Institute for Theoretical Physics, Stanford University,  Stanford, California 94305, USA}

\author{Nicholas P. Butch}
\affiliation{NIST Center for Neutron Research, National Institute of Standards and Technology, Gaithersburg, MD 20899, USA}
\affiliation{Department of Physics, Quantum Materials Center, University of Maryland, College Park, MD 20742, USA}

\date{\today}

\begin{abstract}
We experimentally determine the bounds of the magnetic-field-induced superconducting and magnetic phases near the crystalline $b$ axis of uranium ditelluride (UTe$_2$). By measuring the magnetoresistance as a function of rotation angle and field strength in magnetic fields as large as 41.5 T, we have studied these boundaries in three dimensions of magnetic field direction. The phase boundaries in all cases obey crystallographic symmetries and no additional symmetries, evidence against any symmetry-breaking quadrupolar or higher magnetic order.  We find that the upper critical field of the zero-field superconducting state is well-described by an anisotropic mass model. In contrast, the angular boundaries of the $b$-axis-oriented field-reentrant superconducting phase are nearly constant as a function of field up to the metamagnetic transition, with anisotropy between the $ab$ and $bc$ planes that is comparable to the angular anisotropy of the metamagnetic transition itself. We discuss the relationship between the observed superconducting boundaries and the underlying $\mathbf{d}$ vector that represents the spin-triplet order parameter. Additionally, we report an unexplained normal-state feature in resistance and track its evolution as a function of field strength and angle.
\end{abstract}

\maketitle

\section{Introduction}

The heavy fermion superconductor UTe$_2$ has a unique and complicated phase diagram under applied magnetic fields. Given its orthorhombic crystal structure, it is unsurprising that the upper critical fields of superconductivity in UTe$_2$ are very different for fields along the three crystallographic axes. However, the behavior of superconductivity for fields along the $b$ axis is unusual even in that context.  For crystals with critical temperature of 1.65 K measured at 0.5 K, $H_{c2}$ is roughly 6 T for fields along $a$ and 8 T for fields along $c$, yet superconductivity persists up to approximately 35 T with applied magnetic field along the crystallographic $b$ axis \cite{Ran2019,Aoki2019,Ran2019a,Knafo2021}.

For fields along $b$, the superconducting state is terminated by a metamagnetic transition into a field-polarized (FP) state \cite{Miyake2019}. As the applied field is tilted away from the $b$ axis, the bounds of the FP state evolve quite differently if the tilt is towards the $a$ axis or the $c$ axis \cite{Ran2019a}, indicating the magnetic anisotropy of the system. 

Given the highly anisotropic response of UTe$_2$ to applied magnetic fields, a natural question is how the $b$-axis superconducting state evolves as the applied field is tilted away from $b$. This has been explored in some detail for fields tilted toward the $a$ axis but almost no data exists for fields tilted towards the $c$ axis \cite{Ran2019a, Knebel2019}. Being able to compare the level of anisotropy for this superconducting state to the other phases of UTe$_2$ may provide some clue as to the underlying mechanism of the superconductivity.

There is a further benefit to investigating field orientations beyond the high-symmetry crystallographic directions. Unlike most other uranium-based superconductors, UTe$_2$ is not ferromagnetic; in fact, at ambient pressure no long-range or local dipolar magnetic order has been detected in UTe$_2$ \cite{Sundar2019, Hutanu2020}. Yet other ordered states may exist, such as quadrupolar order, that are more difficult to detect. Beyond obeying the orthorhombic symmetry of the crystal structure,the superconducting phase boundaries could exhibit additional symmetries due to underlying ordered states; these additional symmetries would be reflected in the superconducting phase boundaries as long as superconductivity and the ordered state had any coupling.

As a further motivation, the superconducting state within the FP phase of UTe$_2$ was first found with fields at a seemingly arbitrary direction of 20-40 degrees off the $b$ axis \cite{Ran2019a}. Given this and our lack of understanding of the mechanism for the field-enhanced $b$-axis superconductivity, we performed a full angular survey to characterize its phase boundaries and determine whether any unexpected behavior could be observed with fields outside of the crystallographic planes.

We do not observe any sharp features in the superconducting boundaries as a function of angle, nor do we observe any additional symmetries beyond the crystallographic symmetries of UTe$_2$.  We find that the ratio of angular extent of superconductivity between the $ab$ and $bc$ planes has a maximum as a function of field, as does the angular width of superconducting transitions; we attribute these to a transition between two distinct superconducting phases. Whereas the lower-field superconducting phase is adequately described by an anisotropic effective mass model, the higher-field superconductivity exhibits boundaries that only evolve subtly as a function of field. 

	\begin{figure}
    	\includegraphics[trim=2cm 2cm 14.5cm 1cm, clip=true,width=0.5\textwidth]{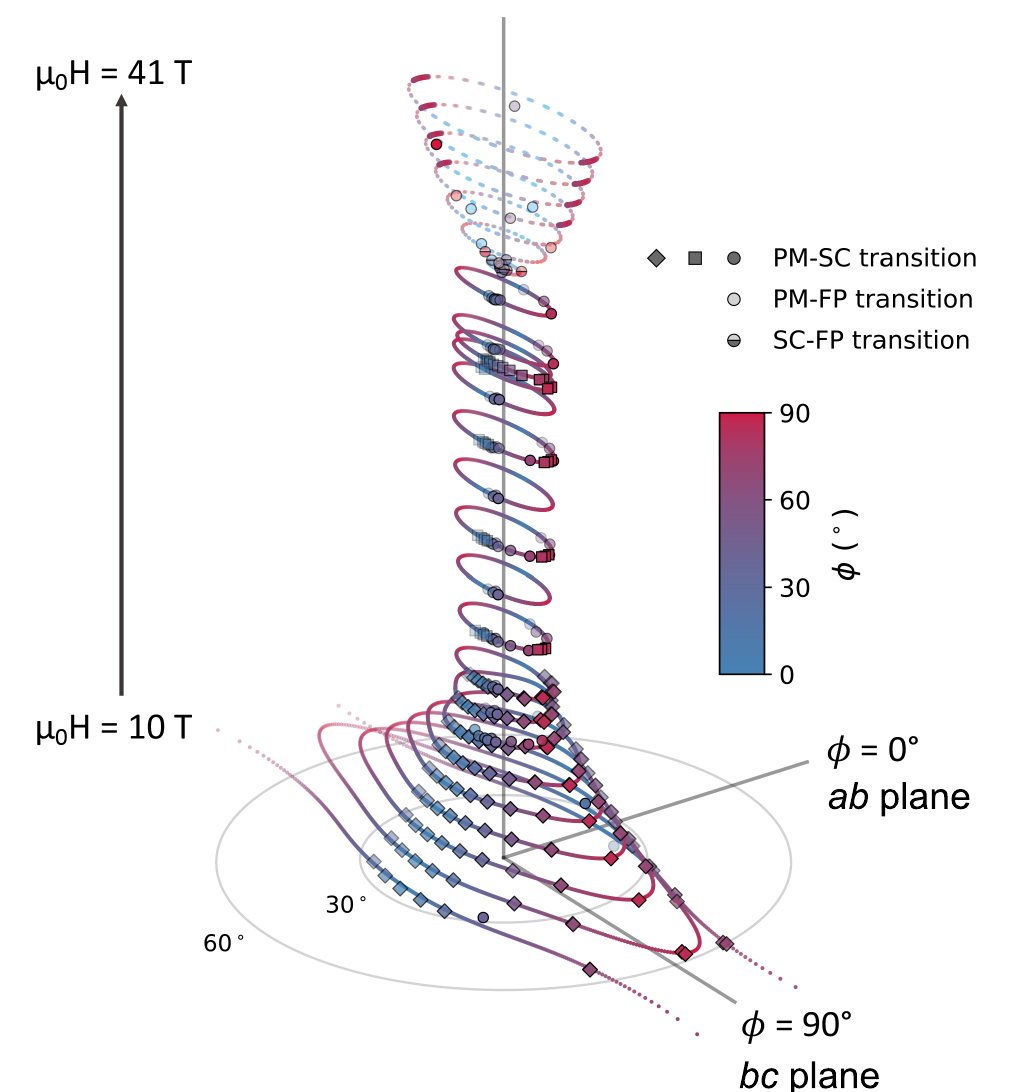}
  	\caption{A phase diagram of UTe$_2$ at approximately 0.4 K as a function of magnetic field strength and direction.  Lines are guides to the eye, with solid lines demarcating the superconducting phase boundaries and dashed lines showing the bounds of the field-polarized state; outside these boundaries the system is paramagnetic.  The different shapes of markers indicate three different datasets gathered on the same sample, in the 18 T magnet (diamonds), 31 T magnet (squares), and 41 T magnet (circles).}
  	\label{fig:fulldiagram}
	\end{figure}

\section{Methods}
\label{sec:methods}

Single crystals of UTe$_2$ used for this study were grown by chemical vapor transport with iodine as a transport agent. Sample 1 was grown using a temperature gradient of 1060/1000 \textdegree C for 1 week, as described in Ref.  \cite{Ran2019}, and has T$_c$ = 1.66 K.  Sample 2 was grown in a 900/830 \textdegree C temperature gradient for 2 weeks and has T$_c$ = 1.89 K. Unless otherwise noted, data shown are from Sample 1, which was measured at the largest density of field strengths and angles.

Magnetoresistance measurements were performed at the National High Magnetic Field Laboratory (NHMFL), Tallahassee; separate datasets were collected using the 18 T superconducting magnet, the 31 T resistive magnet, and the 41 T resistive magnet (all with $^{3}$He inserts) and the 28 T superconducting magnet (with dilution refrigerator).  Crystal orientations were determined using x-ray diffraction. Crystals being measured were mounted on a two-axis rotator, allowing for three-dimensional rotation of the crystal orientation with respect to the applied magnetic field.  Data in this paper were taken at approximately 0.4 K unless otherwise indicated.

\begin{figure}
    	\includegraphics[trim=0cm 1cm 1.5cm 0.5cm, clip=true,width=0.5\textwidth]{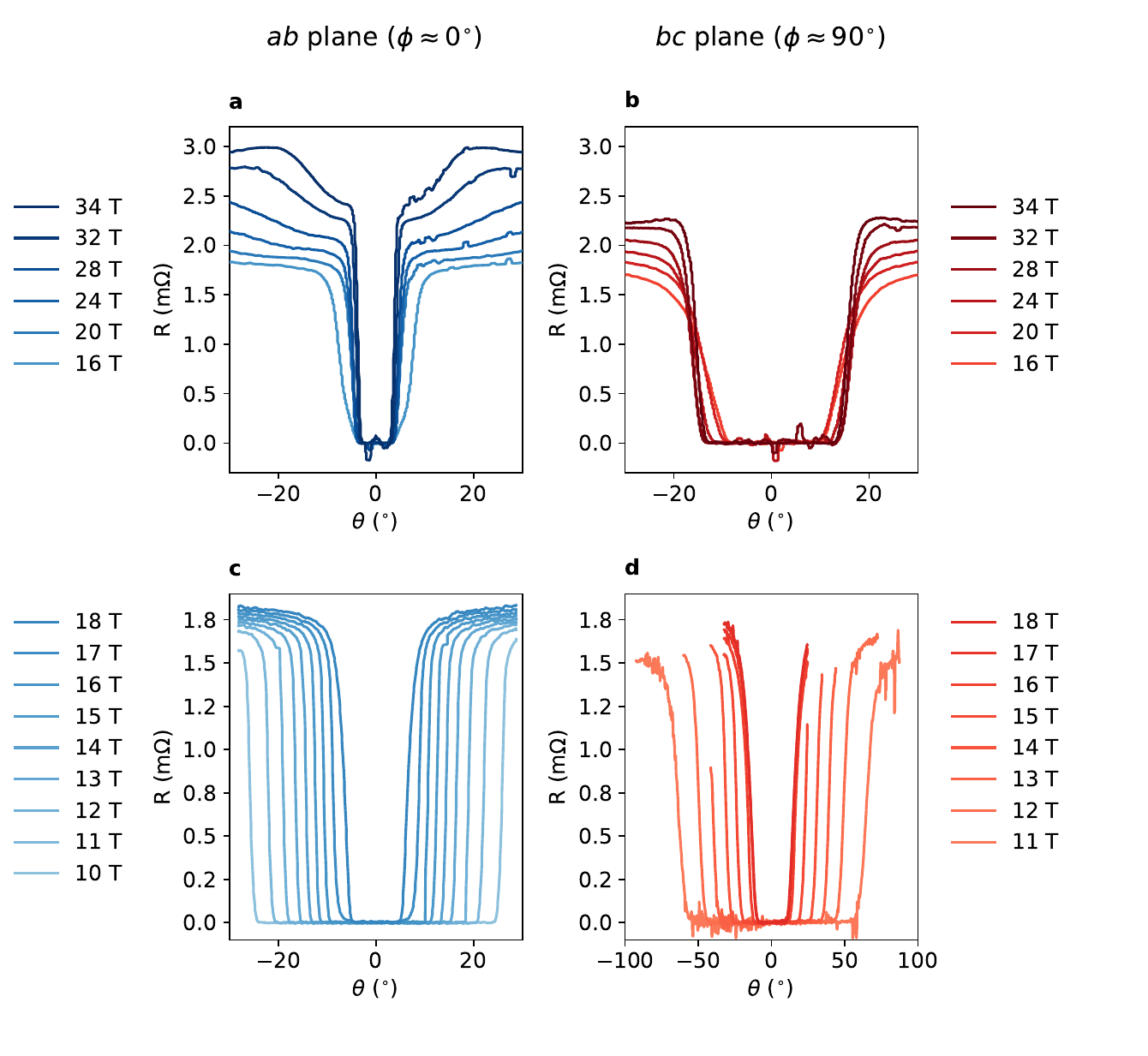}
  	\caption{Representative resistance data as a function of polar angle for fixed magnetic fields and fixed azimuthal angle. Data shown are from Sample 1, approximately in the crystallographic $ab$ plane (a,c) and $bc$ plane (b,d).}
  	\label{fig:Rvsth}
	\end{figure}

Given the extremely anisotropic response of UTe$_2$ to applied magnetic fields, crystal orientation is of utmost importance in creating an accurate phase diagram. We use polar coordinates to describe the magnetic field direction, defining $\theta$ to be the angle of the magnetic field from the $b$ axis and $\phi$ to be its angle from the $a$ axis within the $ac$ plane. Orientation in $\phi$ of our data was confirmed by the expected two-fold rotational symmetry of UTe$_2$. We assured accurate measurement of the boundaries in $\theta$ by always measuring the entire superconducting pocket while rotating $\theta$ and taking the center of the pocket to be at the $b$ axis. Examples of $\theta$-centered data are shown in Fig. \ref{fig:Rvsth}, which shows data taken at fixed fields in the $ab$ and $bc$ planes.

We define the field angle of the superconducting transition as the angle at which $dR/d\theta$ is maximized. Using this definition, we can plot the bounds of superconductivity in $\theta$ as a function of both field magnitude and azimuthal angle $\phi$.

Similarly, for each angular sweep we define the onset and termination of the superconducting transition as the furthest angles from the transition at which $dR/d\theta$ is at least 10 percent of its value at the transition. 

\begin{figure}
    	\includegraphics[trim=1.0cm 1.0cm 0.5cm 2.0cm, clip=true,width=0.5\textwidth]{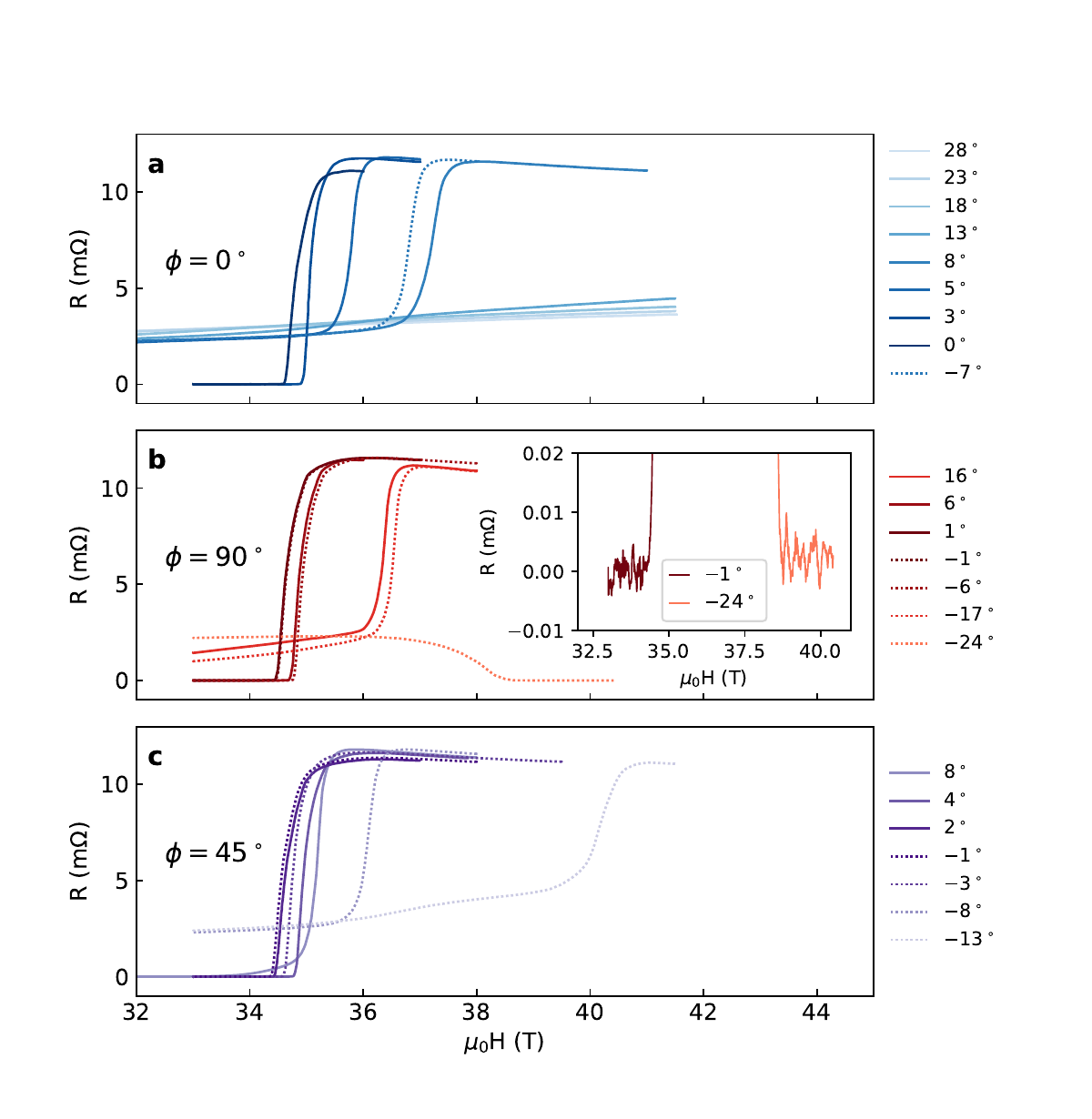}
  	\caption{Resistance as a function of field strength at fixed angles for the determination of the FP phase boundary: (a) for fields at various angles within the the $ab$ plane; (b) for fields at various angles within the $bc$ plane; (c) for fields at various angles within the diagonal plane between the $ab$ and $bc$ planes.  The inset of (b) highlights a field sweep for which the sample entered the field-polarized superconducting state at high fields, in comparison to a field sweep near the $b$ axis with superconductivity only at fields below the metamagnetic transition.  All data plotted were taken with decreasing fields.}
  	\label{fig:RvsH}
	\end{figure}

In addition to the superconducting phase boundaries, we also measured the transition into the FP state in order to compare the anisotropy of these phases.  To measure these transitions we used field sweeps rather than angle sweeps; this was chosen because of the strong torque that is encountered upon crossing into the FP phase, which could hinder angular sweeps.  We gathered data in the $ab$ plane, the $bc$ plane, and midway between the two at $\phi = 45^{\circ}$ as shown in Fig. \ref{fig:RvsH}. The data were taken at known relative values of $\theta$ in each plane; the absolute $\theta$ at which each measurement was taken was determined post-fact by the required symmetry of measurements at positive and negative $\theta$.Similarly to the superconducting boundaries, we defined the edge of the field-polarized phase as the field at which $dR/dH$ is maximized, as there is a large jump in resistance at the metamagnetic transition. Note that one of our measurements also captured the transition into the field-polarized superconducting state of the sample as highlighted in the inset of Fig. \ref{fig:RvsH}.  All of the data shown in Fig. \ref{fig:RvsH} were taken with decreasing fields; see Appendix \ref{app:hysteresis} for a discussion of hysteresis of the metamagnetic transition.

The overall phase diagram that we found using these definitions for the phase boundaries is shown in Fig. \ref {fig:fulldiagram}.

\section{Results and Analysis}
\label{sec:analysis}

\subsection{Evidence for a transition between two superconducting phases}

To date, all nuclear magnetic resonance (NMR) measurements of the Knight shift in UTe$_2$ indicate that it has a spin-triplet superconducting phase \cite{Ran2019,Nakamine2019,Nakamine2021a,Fujibayashi2022}. Therefore, it is reasonable to assume that the main mechanism for external magnetic fields to inhibit superconductivity will be through orbital pair breaking.

The Ginzburg-Landau equation that relates the orbital-limited upper critical field and effective mass for an isotropic superconductor can be extended to a three-dimensional anisotropic material such as UTe$_2$, yielding the following \cite{Takanaka1982}:
\begin{equation}\label{eq:anisomass}
\begin{split}
H_{c2}  =  &\Biggl( \left(\frac{\sin(\theta)\cos(\phi)}{H_{c2a}}\right)^2  + \left( \frac{\sin(\theta)\sin(\phi)}{H_{c2c}}\right)^2  \\
&+ \left(\frac{\cos(\theta)}{H_{c2b}}\right)^2 \Biggr) ^{-1/2},
\end{split}
\end{equation}
where $H_{c2n}$ is the upper critical field with field along the $n$ axis, for $n$ = $a$,$b$,$c$ (see Appendix \ref{app:ratios}).

	\begin{figure*}
    	\includegraphics[trim=5cm 4.2cm 0.5cm 2.5cm, clip=true,width=\textwidth]{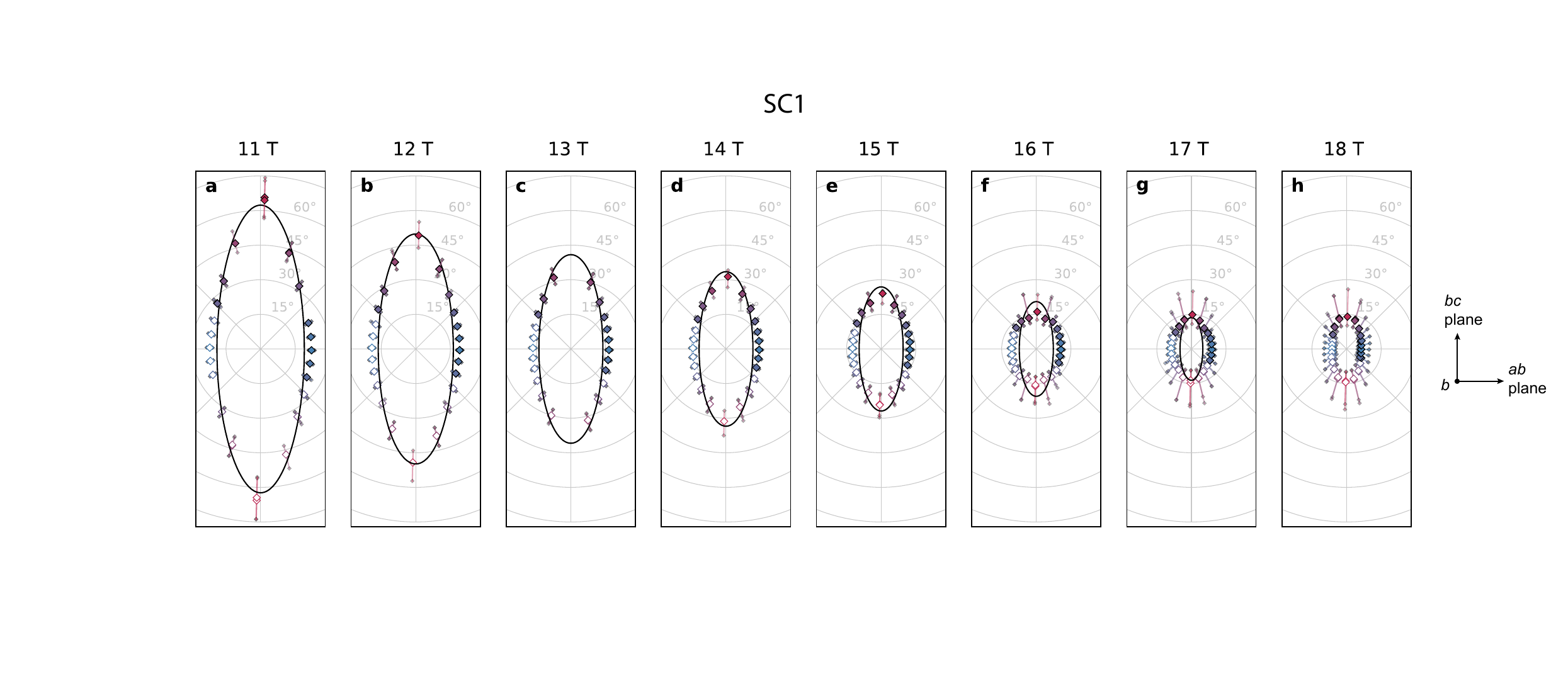}
  	\caption{The bounds of the superconducting phase (large diamonds) as well as the onset and termination of the superconducting transition (small diamonds) are shown as a function of angle at various fixed fields 10 T - 18 T (a-h). In each polar plot the origin represents the $b$ axis, the angular coordinate represents $\phi$,  and the radial coordinate represents $\theta$. Only one marker from each angular sweep is filled; since we have used symmetry to center each angular sweep, the two sides by necessity give the same information. Solid lines in plots (a-g) show a single fit to the anisotropic effective mass model, as described in the text.}
  	\label{fig:anisofits}
	\end{figure*}

Our measurements were taken at fixed field and fixed $\phi$, sweeping $\theta$ to find the edges of superconductivity. By setting $H_{c2} $ equal to the applied field for each data set and using the measured $\theta$ values at which superconductivity is suppressed, we can fit our data to Eq. \ref{eq:anisomass}, using the $H_{c2n}$ values as free parameters.

We begin our analysis with a fit to Eq. \ref{eq:anisomass} using only data from a single constant field strength, using the lowest field possible so as to capture the behavior of the low-field superconducting state. At 10 T, there is no superconducting boundary as a function of $\theta$ in the $bc$ plane, since 10 T is below both $H_{c2b}$ and $H_{c2c}$ for the temperatures at which we measured. Using the 11 T data, we find that the difference between data and fit is minimized with the following values:  $H_{c2a} \approx 4.4$ T, $H_{c2b} \approx 18.0$ T, $H_{c2c} \approx 10.2$  T. The values for $H_{c2a}$ and $H_{c2c}$  are consistent with measured upper critical field values of $H_{c2a} \approx 5$ T and $H_{c2c} \approx 8$ T at 0.5 K for similar CVT-grown samples \cite{Ran2019, Aoki2019}.

The value of $H_{c2b}$ that best fits the data suggests that there are two distinct superconducting phases of UTe$_2$ at ambient pressure, and that at 0.4 K the transition between these phases occurs at roughly 18 T, when the upper critical field of the lower-field phase is reached. This is consistent with thermodynamic evidence for a transition between two superconducting phases: from features in specific heat, Rosuel \textit{et al.} have constructed a field-temperature phase diagram for the lower-field (``SC1") and higher-field (``SC2") states \cite{Rosuel2023}.  Based on this phase diagram, at 0.4 K the SC1-SC2 transition should occur with a field of approximately 18-19 T along the $b$ axis of UTe$_2$. Subtle features in AC susceptibility also indicate a transition between SC1 and SC2; the authors of that study additionally propose an intermediate superconducting phase between SC1 and SC2 in a region of the phase diagram characterized by a low critical current \cite{Sakai2023, Tokiwa2023}. If the SC1-SC2 transition is second-order, such an intermediate phase would be required by thermodynamic considerations \cite{Yip1991}. However, if the SC1-SC2 transition is first-order then no intermediate phase is required to exist.

\begin{figure*}
    	\includegraphics[trim=3cm 0cm 3.3cm 1cm, clip=true,width=\textwidth]{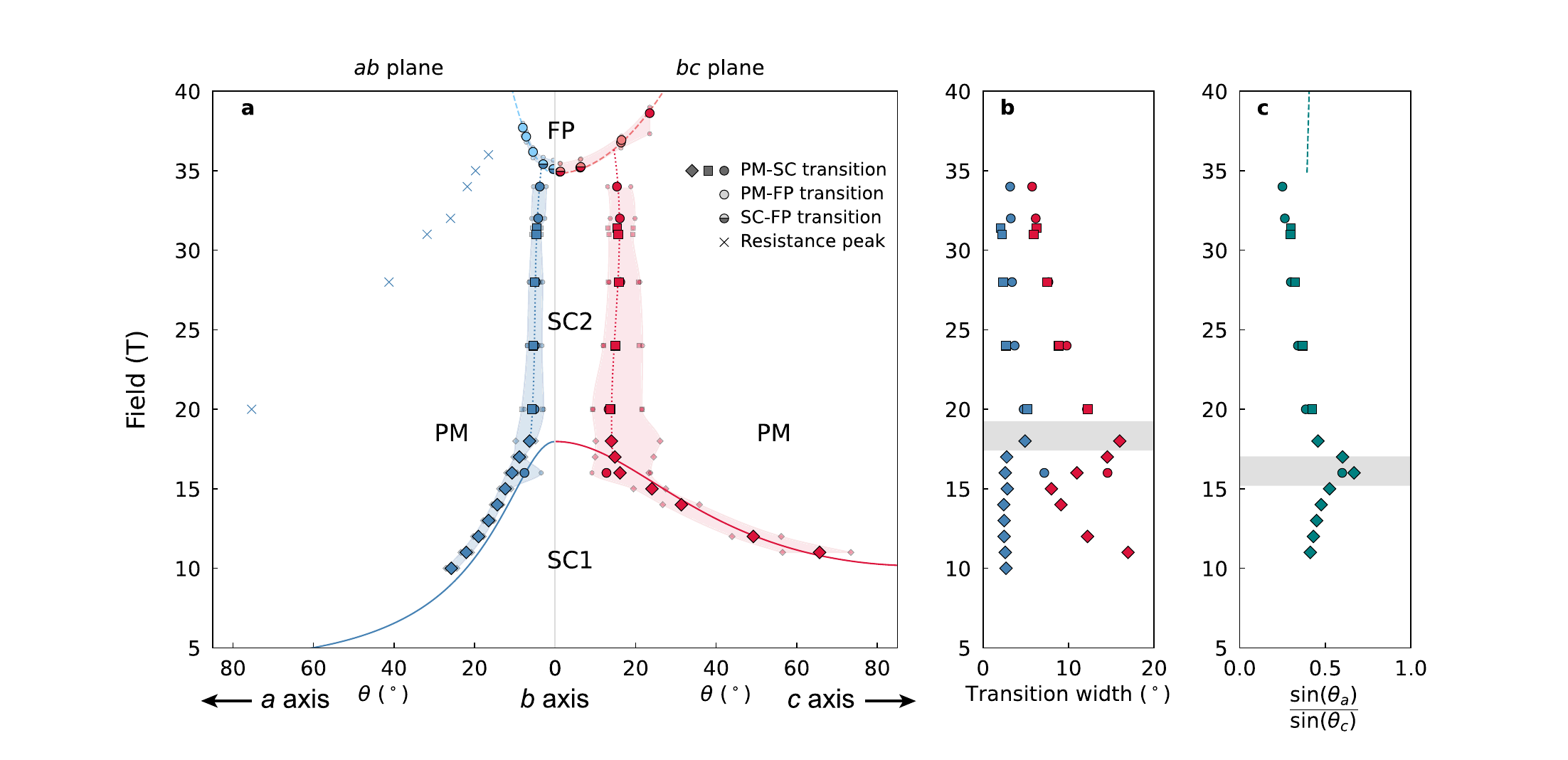}
  	\caption{(a) The phase diagram of UTe$_2$ as a function of field strength and angle for the $ab$ and $bc$ planes.  The widths of transitions are shown by the shaded regions. Solid lines show a fit to the anisotropic effective mass model as described in the text; dashed lines show a paraboloid fit to the FP phase boundaries; dotted lines are guides to the eye.  The resistance peak marked by x's is discussed in Sec.  \ref{subsec:normal} in the text. (b) Angular transition widths for the superconducting phase in the $ab$ plane (blue) and $bc$ plane (red). (c) The ratio $ \nicefrac{\text{sin}(\theta_a)}{\text{sin}(\theta_c)}$, as described in the text.  Shaded regions in (b) and (c) emphasize regions of interest in these plots, as discussed in the text.}
  	\label{fig:planes}
	\end{figure*}

Fig. \ref{fig:anisofits} shows how the anisotropic effective mass model compares to the data when using the values of  $H_{c2n}$ from fits to the 11 T data as described above. The small markers indicate the onset and termination of the superconducting transition, as defined in Sec. \ref{sec:methods}. For the 18 T data no fit is shown, since according to the effective mass model fits there should no longer be superconductivity from SC1 at this field. From the comparison between data and fit between 11 T and 17 T, we can see that the anisotropic effective mass model is a reasonable model for the upper critical fields of UTe$_2$ as a function of field angle and field strength up to approximately 15 T. There are minor features of the data that the effective mass model does not fully capture even in this field range, most notably a heightened extent of superconductivity near the $ab$ plane; we will discuss these discrepancies in Sec. \ref{subsec:SC1}. There is a more marked deviation from the model for data above 15 T,  which we relate to the onset of the SC2 phase at these fields as shown in Fig.  \ref{fig:planes}(a).

The extent of superconductivity in the crystallographic $ab$ and $bc$ planes is shown in Fig. \ref{fig:planes}(a); the best fit to the low-field data using Eq. \ref{eq:anisomass}--the same fit used in Fig. \ref{fig:anisofits}--is shown as solid lines on that phase diagram. The onset and termination of the superconducting phase transition, as defined in Sec. \ref{sec:methods}, are represented by the small markers in Fig. \ref{fig:planes}(a), so the shaded areas between them represent the angular range of the superconducting transition. Taking the difference in angle between onset and termination yields a transition width in degrees, which we show as a function of field in Fig. \ref{fig:planes}(b).

The width of the transition in boths planes is peaked around 18 T, highlighted by the shaded area in Fig. \ref{fig:planes}(b). We propose that the angular broadening of the superconducting transition is related to the transition between the SC1 and SC2 phases.Looking at the fit to the anisotropic effective mass model, we can see that the broadened transition widths occur as the field goes above the expected SC1 phase boundary. For the $bc$ plane, the angular transition width also increases below 14 T, but from Fig. \ref{fig:planes}(a) it appears that this is simply a geometric consequence of the increasing critical angle and the roughly constant transition field width at these angles. If the SC1-SC2 transition of UTe$_2$ is first-order, sample inhomogeneity will lead to a broadened transition in temperature that is most pronounced near the bicritical point  \cite{Yu2022}. A similar mechanism may be responsible for the broadened transition width in terms of the field angle. 

We define $\theta_a$ ($\theta_c$) as the superconducting transition angle in the $ab$ ($bc$) plane. Fig. \ref{fig:planes}(c) shows the ratio $\nicefrac{\text{sin}(\theta_a)}{\text{sin}(\theta_c)}$ for the different field strengths at which we took data.

For a superconductor that is described by Eq. \ref{eq:anisomass}, the ratio $ \nicefrac{\text{sin}(\theta_a)}{\text{sin}(\theta_c)}$ will be a constant (see Appendix \ref{app:ratios} for derivation). We can see that in our data, this is not the case even at low fields, again indicating that even though the low-field superconducting bounds of UTe$_2$ can be decently approximated by Eq. \ref{eq:anisomass}, this is not a complete description of the physics of this system.

Interestingly, while $ \nicefrac{\text{sin}(\theta_a)}{\text{sin}(\theta_c)}$ is less than one for all measured fields, there is a clear maximum in $ \nicefrac{\text{sin}(\theta_a)}{\text{sin}(\theta_c)}$ ratio at 16 T; in other words,this is the measured field strength for which the superconductivity is closest to isotropic in the $ab$ and $bc$ planes, which can also be seen in Fig. \ref{fig:anisofits}.  This trend in $ \nicefrac{\text{sin}(\theta_a)}{\text{sin}(\theta_c)}$ seems to be followed for all of the applicable published data; the exception is a dataset taken at 1 K, a temperature at which the onset of SC2 should occur at higher fields than the termination of SC1 (see Fig. \ref{fig:ratiocomparison}). We therefore speculate that the maximum in $ \nicefrac{\text{sin}(\theta_a)}{\text{sin}(\theta_c)}$ is related to the transition between SC1 and SC2.  We note that 16 T is the field at which the limits of superconductivity in the $bc$ plane begin to sharply deviate from the predictions of the anisotropic effective mass model.  It is interesting that even in the $bc$ plane the transition widths are peaked at a slightly higher field, around 18 T, which is where we approximate the SC1 phase meets the SC2 phase for fields along the $b$ axis.

\subsection{The FP phase}
\label{subsec:FP}
The metamagnetic transition into the FP state of UTe$_2$ occurs at approximately 35 T for field along the $b$ axis. The transition moves to higher fields as the applied field is tilted away from $b$, but the change in transition field with angle is much steeper when tilting towards $a$ than it is when tilting towards $c$ \cite{Ran2019a}. Our own measurements of the FP phase boundaries in the crystallographic planes are shown in Fig. \ref{fig:planes}(a). 

We find that these points, along with phase boundaries extracted from the $\phi = 45^{\circ}$ data shown in Fig. \ref{fig:RvsH}, are well fit by a paraboloid. The dashed lines in Fig. \ref{fig:planes}(a) show this paraboloid fit.  As we measured these boundaries with field sweeps rather than angle sweeps, we do not have transition data at identical fields in the $ab$ and $bc$ planes; we therefore use the paraboloid fit to calculate the ratio $\nicefrac{\text{sin}(\theta_a)}{\text{sin}(\theta_c)}$ for the FP phase boundaries, which is shown as a dashed line in Fig. \ref{fig:planes}(c).  This ratio, which can serve as a metric of anisotropy, is comparable between the SC2 and FP phases, although there is a clear jump at the phase boundary. 

The boundaries of the FP phase are related directly to the magnetic anisotropy of UTe$_2$. The similarity in anisotropy of the FP and SC2 phases indicates that the bounds of the SC2 phase may also depend on magnetic anisotropy, though not directly.  As we will discuss in Sec. \ref{subsec:SC2}, the angle-dependence of the SC2 phase boundaries may be driven by Pauli paramagnetic limiting, if the superconducting $\mathbf{d}$ vector is pinned along the $b$ axis.  This could explain the connection, as the Pauli limiting field should generally depend on the normal-state spin susceptibility \cite{Clogston1962}.

\subsection{The SC1 phase}
\label{subsec:SC1}

The anisotropic effective mass model gives a good first-order description of the evolution of the SC1 phase. However, it fails to capture some details of the angular extent of superconductivity. 

As discussed above, the measured ratio $ \nicefrac{\text{sin}(\theta_a)}{\text{sin}(\theta_c)}$ is not a constant, while it should be constant for a system that is strictly described by Eq. \ref{eq:anisomass}. We can also see directly in Fig. \ref{fig:anisofits} that for all of the measured fields in the SC1 phase, the extent of superconductivity in and near the $ab$ plane is slightly greater than predicted by Eq. \ref{eq:anisomass}.

Previous measurements of the upper critical fields of UTe$_2$ within the crystallographic planes have also indicated deviations from the simple anisotropic effective mass model of Eq. \ref{eq:anisomass}.  One such feature of measurements in the $ab$ plane is a slight local maximum of $H_{c2}$ with field along the $a$ axis \cite{Aoki2019}. It was also noted that at approximately 1 K, superconductivity persists to higher fields near the $b$ axis than the orbital limiting model would predict \cite{Aoki2019}.  At 1 K, there should be a distinct separation between the SC1 and SC2 phases as a function of magnetic field \cite{Knafo2021}. Therefore, this departure from the effective mass model is intrinsic to the SC1 phase. 

One refinement we can make in our model is to consider the role of Pauli paramagnetism, as the model given in Eq. \ref{eq:anisomass} accounts only for orbital pair-breaking effects.   For a spin-triplet superconductor, paramagnetic pair-breaking effects will only be relevant if there are components of the superconducting $\mathbf{d}$ vector along the magnetic field direction \cite{Sigrist2005}.  Therefore, the effects of Pauli paramagnetism will be highly dependent upon the direction of the magnetic field, with this anisotropy arising from the spin-triplet order parameter. Taking this into account, a free energy that incorporates both orbital and paramagnetic limiting can be written for this system. 

\begin{figure}
    	\includegraphics[trim=0.2cm 0cm 0.5cm 0.5cm, clip=true,width=0.5\textwidth]{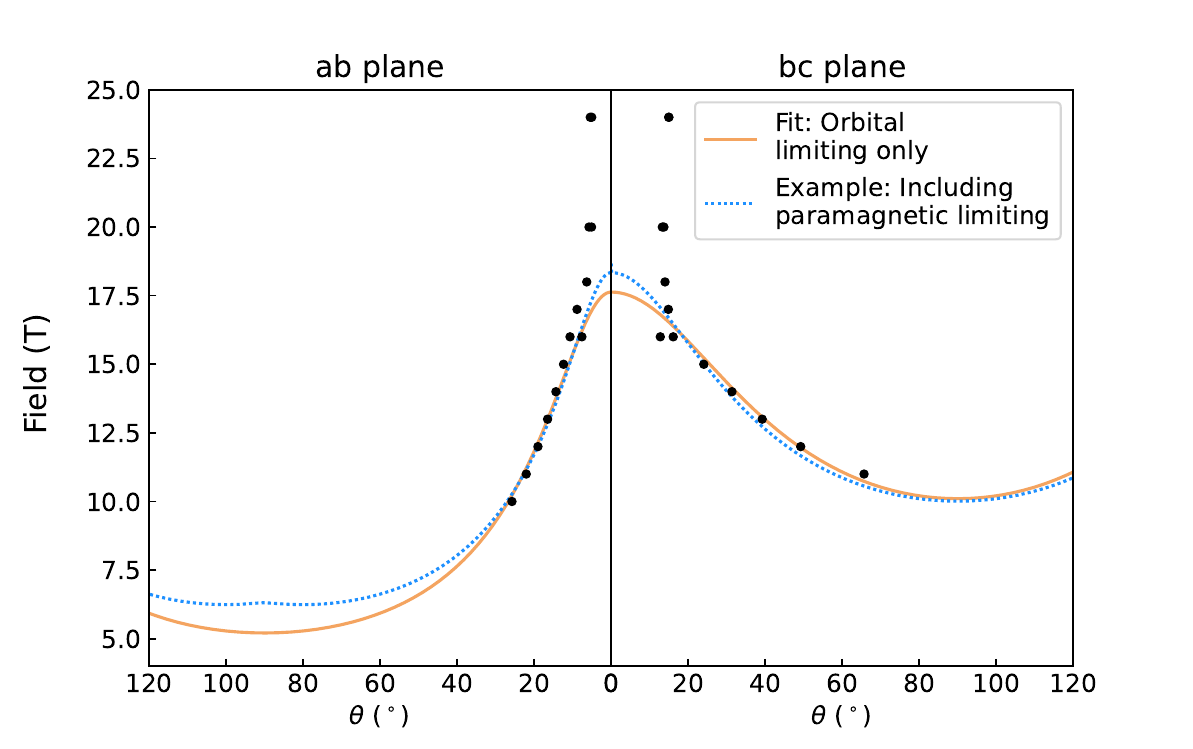}
  	\caption{The measured angular dependence of $H_{c2}$ in the $ab$ and $bc$ planes.  The curve labeled ``Fit: Orbital limiting only" is the best fit to Eq. \ref {eq:TakanakaKjj} using particle swarm optimization followed by least-squares optimization. The curve labeled ``Example: Including paramagnetic limiting" is a plot of Eqs. \ref{eq:abplanemodel} and \ref{eq:bcplanemodel} using parameters chosen to yield a local maximum along the $a$ axis.  The parameters used for both curves are given in Table \ref{table:params}.}
  	\label{fig:Ourfits}
	\end{figure}

As described in Appendix \ref{app:SC1model}, such a model can be used to find analytical expressions for the upper critical field of UTe$_2$ for magnetic fields within the crystallographic $ab$, $ac$, and $bc$ planes.  A fit of these expressions to our own data sets is under-constrained, since the majority of our data were not taken with magnetic field in the crystallographic axes.  We can choose parameters for these expressions such that the upper critical field matches our measured values and has a local maximum along $a$ in the $ab$ plane; an example using such parameters is shown in Fig.  \ref{fig:Ourfits}. Our analytical expressions for $H_{c2}$ can be well-fit to the detailed measurements of upper critical field within the crystallographic planes found in Ref.  \cite{Aoki2019}, as shown in Appendix \ref{app:SC1model}.  The parameters used for both our data and that of Ref.  \cite{Aoki2019} are given in Table \ref{table:params}.

While the model can fit the data well,  it requires high anisotropy of the superconducting coherence lengths in order to yield a local maximum of the upper critical field near the $a$ axis. The parameters from Table \ref{table:params} would require the coherence length along the $a$ axis to be two to three orders of magnitude larger than the coherence lengths along the $b$ and $c$ axes.  Based on the slope of the critical fields near the critical temperature, it has been deduced that the superconducting coherence length is indeed largest along $a$ but that all three coherence lengths are of the same order of magnitude \cite{Paulsen}.

Our analytic model is simple and assumes a fixed $\mathbf{d}$ vector that is independent of magnetic field.  However, as discussed further in Section \ref{subsec:SC2},  NMR measurements indicate that the $\mathbf{d}$ vector of UTe$_2$ will rotate in strong enough applied fields. It is possible that a model taking $\mathbf{d}$ vector rotation into account could fit the data without requiring such strong anisotropy of the superconducting coherence length.  An alternate theory, discussed by Rosuel \textit{et al.}, is that conventional expressions for orbital limiting and paramagnetic limiting are insufficient to describe the upper critical fields of the SC1 phase and that the superconducting pairing itself is field-strength-dependent \cite{Rosuel2023}.

\subsection{The SC2 phase}
\label{subsec:SC2}

	\begin{figure*}
    	\includegraphics[trim=1cm 3cm 0.5cm 0cm, clip=true,width=\textwidth]{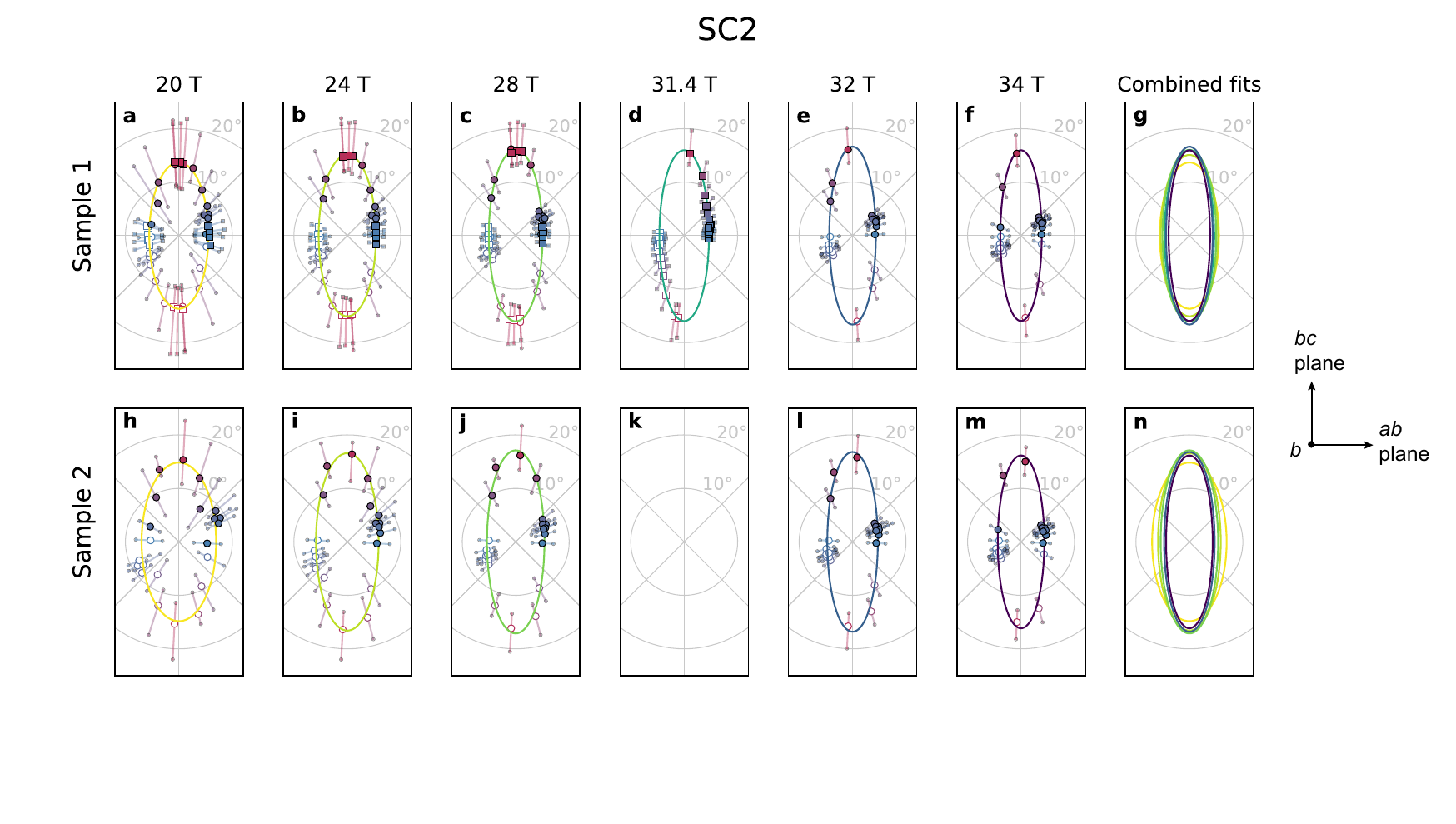}
  	\caption{The bounds of the superconducting phase (large markers) as well as the onset and termination of the superconducting transition (small markers) are shown as a function of angle at various fixed fields 20 T - 34 T for Sample 1 (a-f) and Sample 2 (h-m).  Each dataset has been fit to an ellipse; the fits at all fields are shown for Sample 1 (g) and for Sample 2 (n) to emphasize how little the bounds of superconductivity change with field in this field range. In each polar plot the origin represents the $b$ axis, the angular coordinate represents $\phi$,  and the radial coordinate represents $\theta$. The different shapes of markers indicate two different datasets gathered in the 31 T magnet (squares) and 41 T magnet (circles).  Only one marker from each angular sweep is filled; since we have used symmetry to center each angular sweep, the two sides by necessity give the same information.}
  	\label{fig:highfieldfits}
	\end{figure*}

Fig. \ref{fig:highfieldfits} shows the angular extent of superconductivity for fixed fields ranging from 20 T to 34 T. Just as in Fig. \ref{fig:anisofits},  each large marker indicates the superconducting transition and the smaller markers indicate the onset and termination of the transition.  At these higher fields, the most notable feature of the superconducting phase boundaries is their smooth and simple evolution between the $ab$ and $bc$ planes.  At each fixed field, the phase boundaries appear elliptical; the best-fit ellipse is shown in each panel of Fig. \ref{fig:highfieldfits}. No remarkable features appear, nor is there any evidence of additional symmetries beyond the orthorhombic crystal symmetry of UTe$_2$.

The right panel of Fig. \ref{fig:highfieldfits} shows a combination of all of the elliptical fits for each sample, giving an illustration of the subtle evolution of the SC2 phase boundaries as a function of field strength. Since the phase boundaries change so little with field, such data would be difficult to capture using field sweeps at fixed angles.

In addition to the two-axis-rotator measurements discussed thus far that were taken at approximately 400 mK, we also measured the bounds of superconductivity for fields within the $bc$ plane at multiple fixed temperatures in a dilution refrigerator. We measured at approximately 900 mK, 550 mK, and 50 mK (see Appendix \ref{app:dilfridge} for information on temperature variation during measurements), using both field sweeps at fixed angles and angular sweeps at fixed magnetic fields. 

\begin{figure}
    	\includegraphics[trim=0.3cm 0.2cm 1cm 0.5cm, clip=true,width=0.5\textwidth]{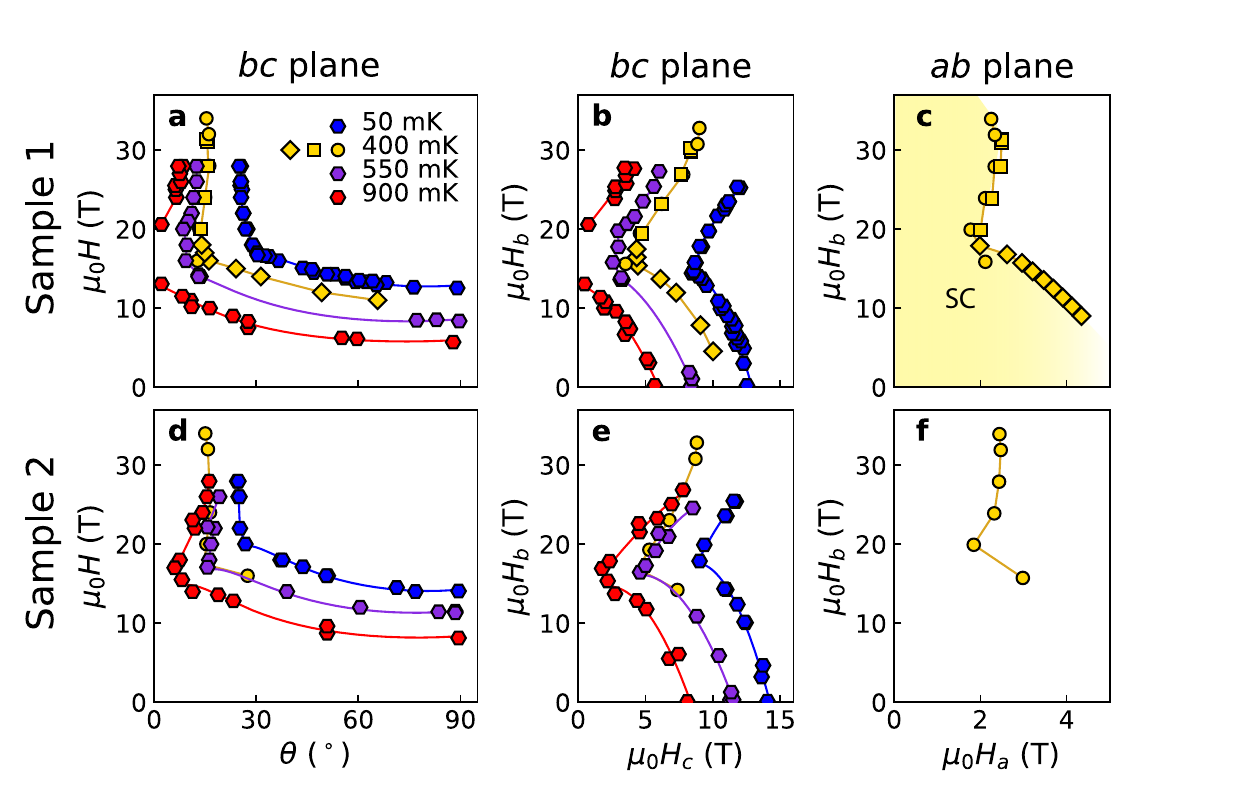}
  	\caption{ The bounds of superconductivity for two different UTe$_2$ samples are shown for various temperatures (a,d) as a function of field strength and field angle within the $bc$ plane; (b,e) as a function of $b$-axis magnetic field and $c$-axis magnetic field; and (c,f) as a function of $b$-axis field and $a$-axis field. All lines are guides to the eye.}
  	\label{fig:dilfridge}
	\end{figure}

The angular extent of superconductivity with respect to field at these temperatures for Sample 1 is shown in Fig. \ref{fig:dilfridge}(a). As expected, the data taken at 550 mK are quite similar to the 400 mK data shown in Fig. \ref{fig:planes}: the angular extent of SC2 slightly increases as the field is increased up to 28 T, the maximum field for these measurements.  At 900 mK there is a clear separation between the SC1 and SC2 phases, consistent with previous reports that for field along the $b$ axis of UTe$_2$, there is a range of temperatures for which there is a normal region between the SC1 and SC2 phases \cite{Knafo2021}.  The 50 mK data show a clear kink in the slope of angle versus field around 17 T,  which we attribute to the SC1-SC2 transition.

In Fig.  \ref{fig:dilfridge}(b), we plot the bounds of superconductivity in terms of the amount of transverse field that destroys superconductivity for a given $b$-axis field.  With the data plotted in this way, we can see that at every temperature we measured the SC2 phase becomes increasingly robust to $c$-axis fields as the $b$-axis field is increased.  This is reminiscent of the field-temperature phase diagram of UTe$_2$: for fields directly along the $b$ axis, the critical temperature of the SC2 phase increases as the applied magnetic field is increased, up until superconductivity abruptly ends at the metamagnetic transition with an applied field of roughly 35 T \cite{Knebel2019, Niu2020, Knafo2021}. 

In contrast, we can see in Fig. \ref{fig:dilfridge}(c) that this monotonic trend is not followed for transverse fields along the $a$ axis: from our 400 mK data, the SC2 phase survives under the highest transverse field for a $b$-axis field of approximately 30 T.  With $b$-axis fields above 30 T, the SC2 state actually becomes less resilient to $a$-axis fields.  The SC2 state is overall less robust to $a$-axis fields than $c$-axis fields, as is clear from its quantitative extent in the $ab$ and $bc$ planes; note the difference in scale of the $x$-axes for Fig.  \ref{fig:dilfridge}(b,e) versus Fig.  \ref{fig:dilfridge}(c,f). But in addition to this, the qualitative evolution of the SC2 phase as a function of $b$-axis field differs in the two planes.  The possible implications of this behavior in regards to the superconducting mechanism and order parameter will be further discussed below.

Sample 2, which has a higher T$_c$ than Sample 1,  exhibits a superconducting region that shrinks more slowly with increasing temperature, as shown in Fig. \ref{fig:dilfridge}(d-f). Even at 900 mK,  Sample 2 does not have a normal state between the SC1 and SC2 phases.  Besides this, the qualitative behavior of Sample 2 is the same as that of Sample 1. \cite{myfootnote}

Based on our analysis above, orbital limiting ends the SC1 phase around 18-19 T at 0.4 K in our samples, so it is natural to ask how the SC2 phase can exist at higher fields.  

The upper critical field due to orbital limiting is proportional to effective mass and there is a good deal of evidence that the effective mass of UTe$_2$ quasiparticles increases with increasing magnetic field along the $b$ axis up to the metamagnetic transition \cite{Knafo2019,Miyake2021a}.  This appears to be due to enhanced longitudinal spin fluctuations that diverge near the metamagnetic transition \cite{Tokunaga2023}. The effective mass does not increase--and in fact decreases--for fields along $a$ and $c$, based on Fermi-liquid fits of magnetoresistance data \cite{Thebault2022}. 

However, enhanced effective mass on its own cannot explain the SC2 phase boundaries. At $\theta = 28^{\circ}$ in the $bc$ plane,  well outside the SC2 region,  UTe$_2$ exhibits a field-dependent increase of effective mass that is comparable in scale to the increase for fields along $b$, and similarly has a maximum at the metamagnetic transition \cite{Miyake2021a}.  So while the high effective mass allows for the existence of the SC2 phase, it does not explain why the SC2 phase is limited to such a small angular region about the $b$ axis.

Next we consider the $\mathbf{d}$ vector that describes the spin-triplet superconducting order parameter. NMR measurements at 0.1 K indicate that for small magnetic fields along the $b$ axis, there is a finite value of $d_b$ \cite{Nakamine2019, Nakamine2021a,Matsumura2023}.  At 7 T along $b$, $d_b$ begins to decrease,  indicating a rotation of the $\mathbf{d}$ vector; once the $b$-axis field is approximately 12.5 T,  $\mathbf{d}$ is entirely perpendicular to the $b$ axis \cite{Nakamine2021}. 

Note that at 0.1 K with 12.5 T along the $b$ axis, the system should still be in the SC1 phase based on our modeling and the phase diagram obtained from specific heat measurements \cite{Rosuel2023}. Moreover,  the $\mathbf{d}$ vector undergoes a similar rotation for fields along the $c$ axis: by 5.5 T along $c$, the $\mathbf{d}$ vector is perpendicular to $c$ \cite{Nakamine2021}, and there is no analog to the SC2 phase for high fields along the $c$ axis.  So the field of the $\mathbf{d}$ vector rotation is not necessarily directly linked to the SC1-SC2 transition. Yet the direction of the $\mathbf{d}$ vector may be a second necessary ingredient for the SC2 phase.

Further measurements have shown that $\mathbf{d}$ remains perpendicular to the $b$ axis for $b$-axis fields up to 24 T, the highest field measured \cite{Kinjo2023}.  It is not known experimentally whether,  in the SC2 phase, the $\mathbf{d}$ vector remains pinned perpendicular to $b$ even as the field tilts slightly away from the $b$ axis or whether the $\mathbf{d}$ vector rotates freely with the applied field.  High-field NMR measurements with fields slightly tilted from the $b$ axis would be instructive in this regard.

If the $\mathbf{d}$ vector is pinned perpendicular to $b$, then paramagnetic limitation would come into play as the field as tilted away from $b$ and could naturally lead to a limited angular range of the SC2 phase about the $b$ axis.  Microscopic calculations of such a scenario have been used to study the transition temperature of the SC2 phase as a function of the field tilt from the $b$ axis \cite{Yu2023}.  It would be interesting for similar calculations to be performed for fixed temperatures to study the resilience of the SC2 phase to transverse $a$-axis and $c$-axis fields and determine whether the qualitative behaviors seen in Fig. \ref{fig:dilfridge} can be reproduced. 

The combination of a field-enhanced effective mass and a $\mathbf{d}$ vector that is pinned perpendicular to $b$ do not elucidate the underlying mechanism or pairing symmetry of the SC2 phase,  but they may account for its high upper critical fields and striking field-angle dependence.  The $\mathbf{d}$ vector being pinned perpendicualr to $b$ would itself be unusual and require further study, given that the $\mathbf{d}$ vector is apparently pinned \textit{along} the $b$ axis for fields below 7 T \cite{Nakamine2021, Nakamine2021a}.

\subsection{Resistive features in the normal state}
\label{subsec:normal}

\begin{figure}
    	\includegraphics[trim=0.8cm 0.2cm 0.5cm 0.5cm, clip=true,width=0.5\textwidth]{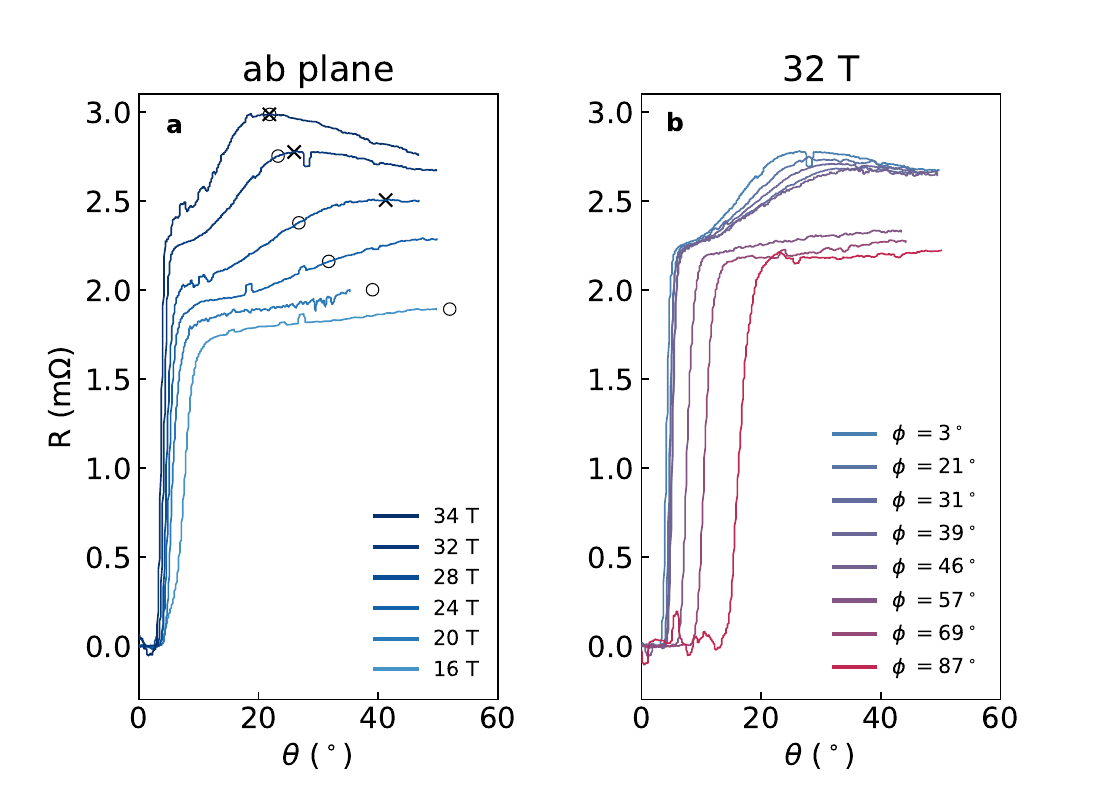}
  	\caption{The hump in resistance as a function of $\theta$ that is seen in the normal state, shown (a) in the $ab$ plane at various field strengths and (b) at 32 T for sweeps taken at various azimuthal angles. In (a), the x markers indicate a maximum in resistance, found after smoothing data. Open circles in (a) represent the predicted evolution of the resistive peak if it occurred at a constant value of the $a$-axis component of the magnetic field, which does not coincide with the observed behavior.}
  	\label{fig:normalstatefeature}
	\end{figure}

While gathering measurements of the superconducting phase boundaries, we also noticed an unexpected feature in the normal-state resistivity of UTe$_2$. As seen in Fig. \ref{fig:Rvsth}(a), there is a hump in the resistance as a function of $\theta$ in the $ab$ plane that is especially noticeable at high fields. Fig \ref{fig:normalstatefeature}(a) shows the evolution of this feature with field strength, while  Fig \ref{fig:normalstatefeature}(b) shows the evolution of the feature at a fixed field but with varying $\phi$. Further plots of the normal-state resistance of Sample 1 can be found in Appendix \ref{app:normalstate}.

At 34 T and with field in the $ab$ plane, this hump is clearly defined and has a maximum around $\theta = 20^{\circ}$. As field is lowered, the maximum in resistivity moves out to higher and higher angles. The amplitude of the hump also steadily decreases as the applied field magnitude is lowered. At 16 T, the lowest field at which we took a broad enough angular sweep to examine the normal state, there appears to still be a subtle kink in the resistivity as a function of $\theta$.

We have extracted the $\theta$ at which the resistive maximum appears for angular sweeps in the $ab$ plane for all of the field strengths at which we have adequate data. The positions of these resistive maxima are plotted in Fig. \ref{fig:planes}(a) along with the superconducting and metamagnetic phase boundaries. 

As we measure at azimuthal angles further from the $ab$ plane, the amplitude of the resistive feature decreases and it moves to higher angles. At 32 T, as shown in Fig. \ref{fig:normalstatefeature}(b), the feature is visible at $\phi = 46^{\circ}$ but has disappeared by $\phi = 57^{\circ}$. We do not believe that the hump has simply moved to a higher $\theta$ than we measured,  as we have full angular scans in the $bc$ plane with fields up to 28 T that do not show evidence of this resistive feature (see Appendix \ref{app:normalstate}).

This feature is not clearly tied to Fermi surface effects. Angle-dependent magnetoresistance oscillations are a tempting explanation, as their amplitude will be larger at higher fields. However, such features arise purely from the Fermi surface geometry, so their position in $\theta$ should not be field-dependent.

A similar resistive maximum was found in magnetoresistance measurements of UCoGe, tentatively attributed to a magnetic transition or Lifshitz transition \cite{Bay2014}. In that material, regardless of field angle the resistive maximum always occurred when the $c$-axis component of the magnetic field reached approximately 8.5 T. Based on the position of our feature at approximately $\theta = 22^{\circ}$ at 34 T in the $ab$ plane of UTe$_2$,  we might surmise that the resistive maximum occurs when the $a$-axis component of magnetic field is roughly 12.6 T. However, that is not consistent with the peak's evolution as a function of field strength, as shown in Fig. \ref{fig:normalstatefeature}(a). The open circles indicate the value of $\theta$ for each angular sweep at which the $a$-axis component of field would be 12.6 T; they do not track the maximum in resistivity.

Given that the hump is most pronounced in the $ab$ plane and that the $a$ axis is the easy magnetic axis of UTe$_2$, it seems likely that the observed maximum in resistance is due to enhanced spin fluctuations,  perhaps signifying field angles at which competing energy scales become comparable.

\section{Conclusions}

We have performed a full survey of the low-temperature boundaries of superconductivity in UTe$_2$ with varying field strengths and directions, for fields up to 35 T; we have also studied the bounds of the field-polarized phase up to 41.5 T.We have found that the bounds of superconductivity evolve smoothly between the crystallographic planes, without signatures of any additional symmetries beyond the orthorhombic symmetry of the crystal. Our modeling indicates that the SC1 and SC2 phases are indeed distinct, with the SC1 phase terminating at approximately 18 T at 0.4 K. This is also indicated by a maximum in angular superconducting transition widths and a maximum in our measure of anisotropy near the SC1-SC2 boundary. We have measured the SC2 phase boundaries in great detail, revealing the subtle evolution of these boundaries with field, and have discussed pinning of the $\mathbf{d}$ vector in the $ac$ plane as a likely cause. In addition, our discovery of a normal-state resistive feature suggests an as-yet unknown magnetic scattering mechanism that may be relevant in understanding the unusual superconductivity of UTe$_2$.

\section{Acknowledgments}
This work was supported in part by the National Science Foundation under the Division of Materials Research Grant NSF-DMR 2105191. A portion of this work was performed at the National High Magnetic Field Laboratory, which is supported by National Science Foundation Cooperative Agreement No. DMR-1644779 and the State of Florida.  Sample preparation and characterization was supported by the Department of Energy Award No. DE-SC-0019154 and the Gordon and Betty Moore Foundation’s EPiQS Initiative through Grant No. GBMF9071.  J.J.Y.~was supported by the National Science Foundation Graduate Research Fellowship under Grant No.~DGE-1656518. The authors declare no competing financial interest.  Identification of commercial equipment does not imply recommendation or endorsement by NIST.

\appendix

\section{The anisotropic effective mass model}
\label{app:ratios}

For an anisotropic superconductor described by the effective mass model, the upper critical field can be written as
\begin{equation}\label{eq:Takanaka}
H_{c2} = \frac{c  \lvert \alpha \rvert}{e \hbar} \left( \frac{n_x^2}{m_2m_3}  + \frac{n_y^2}{m_3m_1}  + \frac{n_z^2}{m_1m_2} \right) ^{-1/2}
\end{equation}
where the applied magnetic field is in the direction of the unit vector $\hat{n}$ and $m_{1,2,3}$ are the effective masses of the quasiparticles along the principal axes $a$, $b$, and $c$; $\alpha$ is a temperature-dependent phenomenological parameter from Ginzburg-Landau theory \cite{Takanaka1982}.

If we define $H_{c2a}$ to be the critical field when $\hat{n} = (1,0,0)$, etc., we can see that Eq. \ref{eq:Takanaka}  is equivalent to

\begin{align}
H_{c2} & =  \bigg( \left(\frac{\sin(\theta)\cos(\phi)}{H_{c2a}}\right)^2  + \left( \frac{\sin(\theta)\sin(\phi)}{H_{c2c}}\right)^2  \\
&+ \left(\frac{\cos(\theta)}{H_{c2b}}\right)^2 \bigg) ^{-1/2} \nonumber
\end{align}
where we have defined $\theta$ to be the angle of the magnetic field from the $b$ axis and $\phi$ to be the angle of the field from the $a$ axis within the $ac$ plane.

In our measurements, we kept field fixed and, at a given $\phi$, swept $\theta$ to find the bounds of superconductivity.

In that sense, taking $H$ and $\phi$ as constants, we are finding the value of $\theta$ that fulfills

\begin{align}
H & =  \bigg( \left(\frac{\sin(\theta)\cos(\phi)}{H_{c2a}}\right)^2  + \left( \frac{\sin(\theta)\sin(\phi)}{H_{c2c}}\right)^2   \\
& + \left(\frac{\cos(\theta)}{H_{c2b}}\right)^2 \bigg) ^{-1/2} \nonumber
\end{align}

Consider field in the $ab$ plane. Take $H$ to be fixed, and take $\theta_a$ to be the angle at which the critical field is equal to the applied field. Then

\begin{equation}
\label{eq:abplane}
H =  \left( \left(\frac{\sin(\theta_a)}{H_{c2a}}\right)^2  + \left(\frac{\cos(\theta_a)}{H_{c2b}}\right)^2 \right) ^{-1/2}.
\end{equation}

This equation has a solution for $\theta_a$ as long as the field at which we are measuring is between $H_{c2a}$ and $H_{c2b}$. Rearranging Eq. \ref{eq:abplane} and employing trigronemtric substitution, we find
\begin{equation}
\label{eq:ab}
\left(  \left(\frac{H}{H_{c2a}}\right)^2 -  \left(\frac{H}{H_{c2b}}\right)^2 \right)\sin^2(\theta_a) + \left(\frac{H}{H_{c2b}}\right)^2 = 1.
\end{equation}

Similarly, for field in the $bc$ plane, at the same field strength $H$, we find

\begin{equation}
\label{eq:bc}
\left(  \left(\frac{H}{H_{c2c}}\right)^2 -  \left(\frac{H}{H_{c2b}}\right)^2 \right)\sin^2(\theta_c) + \left(\frac{H}{H_{c2b}}\right)^2 = 1.
\end{equation}

Combining Eq. \ref{eq:ab} and Eq. \ref{eq:ab}, we can find the ratio 
\begin{equation}
\frac{\sin(\theta_a)}{\sin(\theta_c)} = \sqrt{\frac{\left(\frac{H_{c2b}}{H_{c2c}}\right)^2 -  1 }{\left(\frac{H_{c2b}}{H_{c2a}}\right)^2 -  1}}.
\end{equation}

If we are measuring each pair of $\theta_a$ and $\theta_c$ at a constant field, and if the anisotropic mass model is valid, the ratio of $\sin(\theta_a)$ to $\sin(\theta_c)$ should not depend at all on the field at which they are measured.

\begin{figure}
    	\includegraphics[trim=0.3cm 1.5cm 0.5cm 0.9cm, clip=true,width=0.5\textwidth]{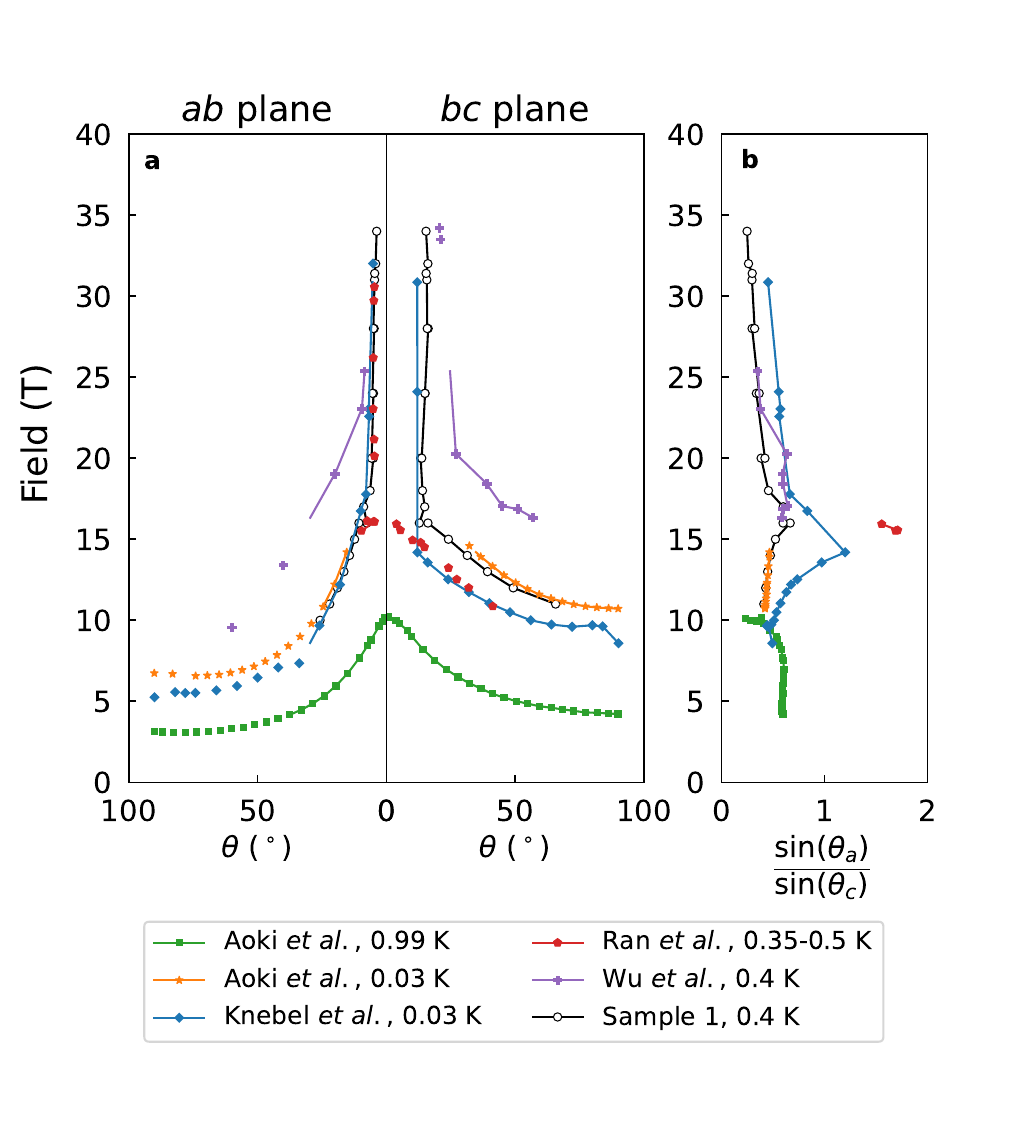}
  	\caption{(a) The bounds of superconductivity of Sample 1 in the $ab$ and $bc$ planes (circles) are shown with similar data from \cite{Aoki2019} (squares and stars), \cite{Knebel2019} (diamonds), \cite{Ran2019a} (pentagons), and \cite{Wu2023} (pluses).  (b) The ratio $\nicefrac{\text{sin}(\theta_a)}{\text{sin}(\theta_c)}$, as described in the text, is shown for all of these datasets for the field ranges with data in both planes.} 
  	\label{fig:ratiocomparison}
	\end{figure}

As shown in Fig. \ref{fig:planes}, this ratio is not constant with field for Sample 1, and in fact has a maximum around 16 T. In Fig. \ref{fig:ratiocomparison} we show similar data taken from published works on UTe$_2$ to see if this trend holds across other measured samples. 

The previously published datasets do not include measurements at identical fields for the $ab$ plane and $bc$ plane. Therefore, in order to calculate the ratio $\nicefrac{\text{sin}(\theta_a)}{\text{sin}(\theta_c)}$, we have linearly extrapolated between the measured datapoints. The lines connecting data in Fig. \ref{fig:ratiocomparison}(a) show these extrapolations; we have only shown lines for the range of fields for which we had (extrapolated) data in both the $ab$ and $bc$ planes. Fig. \ref{fig:ratiocomparison}(b) shows the ratios calculated in this manner; we calculated the ratio for every field at which there was a measured data point in the $ab$ plane and/or the $bc$ plane.

Most of the datasets we studied do not allow for calculation of $\nicefrac{\text{sin}(\theta_a)}{\text{sin}(\theta_c)}$ across a wide field range,  but still seem consistent with our data. The broadest dataset, from Ref. \cite{Knebel2019}, shows an incredibly similar trend to ours: a ratio that steadily increases with increasing field up to a maximum (in that case near 14 T) and then steadily decreases. The dataset taken near 1 K does not show this trend, as discussed in the main text. Note that the actual value of the ratio $\nicefrac{\text{sin}(\theta_a)}{\text{sin}(\theta_c)}$ may not be accurate if there was sample misalignment for the measurements shown, but the trend in the ratio with field should be fairly robust.

\section{Analytical model for SC1}
\label{app:SC1model}

For our model, we assume a single-band, single-order-parameter model and begin with Ginzburg and Landau's typical expression for the free energy of a superconductor:

\begin{align}
\label{eq:energy1}
F_s - F_n & = \int d^3x \bigg(\alpha(T)|\psi|^2 + \beta(T) |\psi|^4\\
& + \sum_{jk} K_{jk} ((\partial_{x_j}- \frac{2ei}{c} A_j)\psi)^* ((\partial_{x_k} -\frac{2ei}{c} A_k)\psi)\bigg) \nonumber
\end{align}
In Eq. \ref{eq:energy1}, $\alpha(T)$ and $\beta(T)$ are the typical phenomenological parameters of Ginzburg-Landau theory; $K$ is a diagonal tensor,  reflecting the orthorhombic symmetry of UTe$_2$ that allows for anisotropic kinetic terms; and $\mathbf{A}$ is the magnetic vector potential.

We then include an induced magnetization term:

\begin{align}
\label{eq:energy2}
F_s - F_n & = \int d^3x \bigg(\alpha(T)|\psi|^2 + \beta(T) |\psi|^4  \\
& + \sum_{jk} K_{jk} ((\partial_{x_j} - \frac{2ei}{c} A_j)\psi)^* ((\partial_{x_k} -\frac{2ei}{c} A_k)\psi)  \nonumber \\ 
&+ \sum_{jk}  H_j H_k \int\frac{d\Omega}{4\pi}\chi_N^{jk} \bigg[1 \nonumber \\ 
&\ \ \ \ \ \ \ \ \ \ \ \ \ \ \ \ \ -(1-Y(\mathbf{n}, T))\frac{d_j^*(\mathbf{n})d_k(\mathbf{n})}{|\mathbf{d}(\mathbf{n})|^2} \bigg] \bigg) \nonumber.
\end{align}
Here, $\chi_N$ is the normal-state susceptibility tensor,  $Y(\mathbf{n}, T)$ is the momentum-dependent Yoshida function, and $\mathbf{d}(\mathbf{n})$ is the vector representing the spin-triplet order parameter. Note that the susceptibility tensor is diagonal for an orthorhombic system such as UTe$_2$.

We take $x$, $y$, $z$ to be along the $a$, $b$, and $c$ axes respectively.  Then an applied magnetic field can be described by $\mathbf{H}= H (\sin\theta \cos\phi, \cos\theta, \sin\theta\sin\phi)$, using the definitions of $\theta$ and $\phi$ from the main text.  If the field lies in a plane defined by the crystal axes, it is simple to solve for $H_{c2}$ by choosing a gauge depending on only one coordinate. 

We make the following assumptions:
\begin{enumerate}
	\item That fourth-order gradient terms in the Ginzburg-Landau free energy can be discounted;
	\item That $|\psi|^2$ vanishes at infinity, which causes certain terms to be 0 when integrated by parts;
	\item That $H$ is close to $H_{c2}$, such that $\psi$ is small.
\end{enumerate}

If the magnetic field is in the $ab$ plane, $\phi$ = 0 and $\mathbf{H}=H(\sin\theta, \cos\theta,0)$. We can choose the gauge $\mathbf{A}=Hz(\cos\theta, -\sin\theta,0)$ to produce such a magnetic field.
We assume that $\psi$ has the structure of a generalized Landau level solution: for field in the $ab$ plane, this means $\psi\sim e^{ik_x x}e^{ik_yy}u(z)$.

We vary Eq.  \ref{eq:energy2} with respect to $\psi^*$ to find the Ginzburg-Landau equation for $\psi$, using the assumptions stated above. This equation ends up imposing a self-consistency condition on $H$; the maximum $H$ that satisfies this condition will be the upper critical field.

For convenience,  we define
\begin{equation}
F(d_i) \equiv \frac{7}{4}\zeta(3)  \int \frac{d\Omega}{4\pi} \frac{|f(\mathbf{n})|^2}{\pi^2k_B^2T^2} \frac{d_i^*(\mathbf{n})d_i(\mathbf{n})}{|\mathbf{d}(\mathbf{n})|^2}.
\end{equation}

For field in the $ab$ plane, the full solution for $H_{c2}$ is: 
\begin{align}
\label{eq:abplanemodel}
H_{c2}(\theta, \phi=0) = \frac{e}{c} \frac{\left( K_{zz}K_{xx}\cos^2\theta  + K_{zz}K_{yy}\sin^2\theta \right)^{1/2}}{\chi_N^{xx} \sin^2\theta F(d_x) + \chi_N^{yy}  \cos^2\theta F(d_y)} \\
\times  \left[  \sqrt{1+\frac{c^2 \vert \alpha \vert [\chi_N^{xx} \sin^2\theta F(d_x) + \chi_N^{yy}  \cos^2\theta F(d_y)]}{e^2\left( K_{zz}K_{xx}\cos^2\theta  + K_{zz}K_{yy}\sin^2\theta \right) }} -1 \right]. \nonumber
\end{align} 

Similarly,  we set $\phi=\pi/2$ to get magnetic fields in the $bc$ plane, for which we can use the gauge $\mathbf{A}=Hx(0,  \sin\theta , -\cos \theta)$. For fields in the $bc$ plane, we find

\begin{align}
\label{eq:bcplanemodel}
H_{c2}(\theta, \phi= \tfrac{\pi}{2}) = \frac{e}{c} \frac{\left( K_{zz}K_{xx}\cos^2\theta  + K_{xx}K_{yy}\sin^2\theta \right)^{1/2}}{\chi_N^{zz} \sin^2\theta F(d_z) + \chi_N^{yy}  \cos^2\theta F(d_y)} \\
\times  \left[  \sqrt{1+\frac{c^2 \vert \alpha \vert  [\chi_N^{zz} \sin^2\theta F(d_z) + \chi_N^{yy}  \cos^2\theta F(d_y)]}{e^2\left( K_{zz}K_{xx}\cos^2\theta  + K_{xx}K_{yy}\sin^2\theta \right) }} -1 \right]. \nonumber
\end{align} 

We set $\theta=\pi/2$ to get magnetic fields in the $ac$ plane, for which we can use the gauge $\mathbf{A}=Hy(-\sin\phi ,  0, \cos \phi)$. For fields in the $ac$ plane, we find
\begin{align}
H_{c2}(\theta = \tfrac{\pi}{2}, \phi) = \frac{e}{c} \frac{\left( K_{zz}K_{yy}\cos^2\phi  + K_{xx}K_{yy}\sin^2\phi \right)^{1/2}}{\chi_N^{zz} \sin^2\phi F(d_z) + \chi_N^{xx}  \cos^2\phi F(d_x)} \\
\times  \left[  \sqrt{1+\frac{c^2 \vert \alpha \vert [\chi_N^{zz} \sin^2\phi F(d_z) + \chi_N^{xx}  \cos^2\phi F(d_x)]}{e^2\left( K_{zz}K_{yy}\cos^2\phi + K_{xx}K_{yy}\sin^2\phi \right) }} -1 \right]. \nonumber
\end{align} 

Using Eq. \ref{eq:abplanemodel} and Eq. \ref{eq:bcplanemodel}, we can perform a simultaneous fit for data from both the $ab$ and $bc$ planes.  Notice that the terms $F(d_j)$ in these equations always appear with a factor of $\chi_N^{jj}$; we define $f_j \equiv F(d_j)\chi_N^{jj}$ for $j = x,y,z$. For ease, we also combine a factor of $\frac{e}{c}$ with each $K_{jj}$.  The free parameters in such a fit are then $f_j$ and $\frac{e}{c}K_{jj}$ for $j = x,y,z$. 

One can show algebraically that the right sides of Eq. \ref{eq:abplanemodel} and Eq. \ref{eq:bcplanemodel} will be unchanged if the Ginzburg-Landau parameter $\alpha$,  all of the $K_{jj}$, and all of the $f_j$ are scaled by some constant. This means that through fits to data, we can not determine the actual values of the parameters $K_{jj}$ or $f_j$, only their relative values. We fix $\vert \alpha \vert = 1$ for the purpose of fitting.

For easier comparison with our model, we can also rewrite Eq.  \ref{eq:Takanaka} in terms of the $K_{jj}$:

\begin{equation}\label{eq:TakanakaKjj}
\begin{split}
H_{c2}  = \frac{c \vert \alpha \vert }{2 e} &\Biggl( K_{yy}K_{zz}\left(\sin(\theta)\cos(\phi)\right)^2  \\
&+ K_{xx}K_{yy}\left( \sin(\theta)\sin(\phi)\right)^2  \\
&+ K_{xx}K_{zz}\left(\cos(\theta)\right)^2 \Biggr) ^{-1/2}.
\end{split}
\end{equation}
This is the expression for the upper critical field with purely orbital limiting.

	\begin{figure}
    	\includegraphics[trim=0.2cm 0.2cm 0.8cm 0.8cm, clip=true,width=0.5\textwidth]{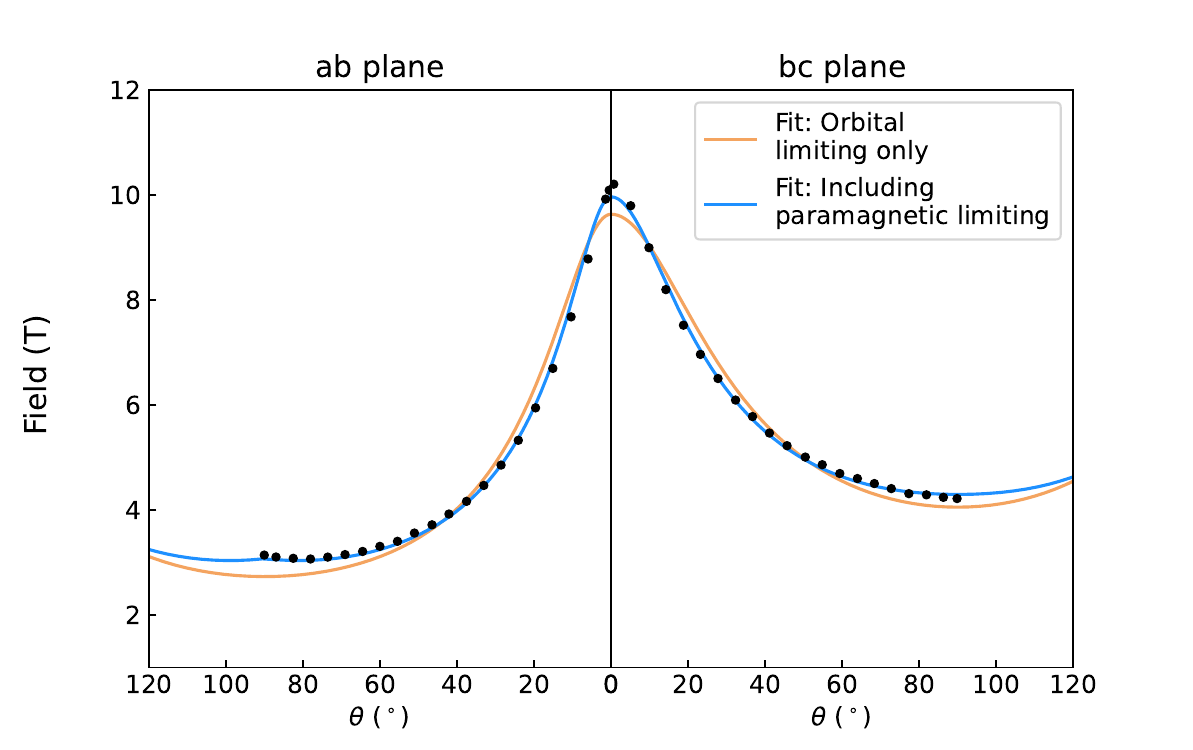}
  	\caption{The angular dependence of $H_{c2}$ in the $ab$ and $bc$ planes at 1 K, from Ref.  \cite{Aoki2019}.  The curve labeled ``Fit: Orbital limiting only" is the best fit to Eq. \ref {eq:TakanakaKjj} while the curve labeled ``Fit: Including paramagnetic limiting" is the best fit using Eqs. \ref{eq:abplanemodel} and \ref{eq:bcplanemodel}.  Fits were carried out using particle swarm optimization followed by least-squares optimization. The parameters obtained from these fits are given in Table \ref{table:params}.}
  	\label{fig:1Kfits}
	\end{figure}

\begin{center}
\begin{table}
\begin{tabular}{|c|c|c|c|c|}
\hline
&\multicolumn{2}{|c|}{}&\multicolumn{2}{|c|}{} \\
&\multicolumn{2}{|c|}{Our data}&\multicolumn{2}{|c|}{Data at 1 K from Ref. \cite{Aoki2019}} \\
&\multicolumn{2}{|c|}{}&\multicolumn{2}{|c|}{} \\
\hline
&&&&\\
& Example: & Fit: & Fit:  & Fit:  \\
& Eqs. \ref{eq:abplanemodel}, \ref{eq:bcplanemodel}  & Eq.  \ref{eq:Takanaka}  & Eqs. \ref{eq:abplanemodel},  \ref{eq:bcplanemodel}  & Eq.  \ref{eq:Takanaka} \\
&&&&\\
\hline
&&&&\\
$\dfrac{F(d_y)\chi_N^{yy}}{F(d_x)\chi_N^{xx}} $ & $\approx$ 0 & - &  $\approx$ 0 & - \\
&&&&\\
\hline
&&&&\\
$\dfrac{F(d_z)\chi_N^{zz}}{F(d_x)\chi_N^{xx}}$ & 0.21 & - & 0.42 & -  \\ 
&&&&\\
\hline
&&&&\\
$\dfrac{K_{yy}}{K_{xx}}$ & 5.5e-4  & 11.5 & 4.1e-6 & 12.4\\ 
&&&&\\
\hline
&&&&\\
$\dfrac{K_{zz}}{K_{xx}}$ & 7.4e-4  & 3.8 & 2.2e-5 & 2.2\\ 
&&&&\\
 \hline
\end{tabular}
\caption{The parameters used to achieve the curves plotted in Fig. \ref{fig:Ourfits} and Fig. \ref{fig:1Kfits}.  Only the relative values of the parameters are shown; as explained in the text, fitting for all parameters was only done up to a constant of proportionality, so the absolute value of each parameter is not meaningful.}
\label{table:params}
\end{table}
\end{center}

The parameters used for Fig. \ref{fig:Ourfits} and Fig.  \ref{fig:1Kfits} are given in Table \ref{table:params}. Note that the superconducting coherence length should be proportional to $\sqrt{K}$. Therefore the ratios of the $K_{jj}$ for our model would require an extremely anisotropic coherence length, as discussed in the main text.

\section{Additional normal-state data: feature in resistance}
\label{app:normalstate}

	\begin{figure*}[t]
    	\includegraphics[trim=3cm 0.5cm 3cm 1.8cm, clip=true,width=\textwidth]{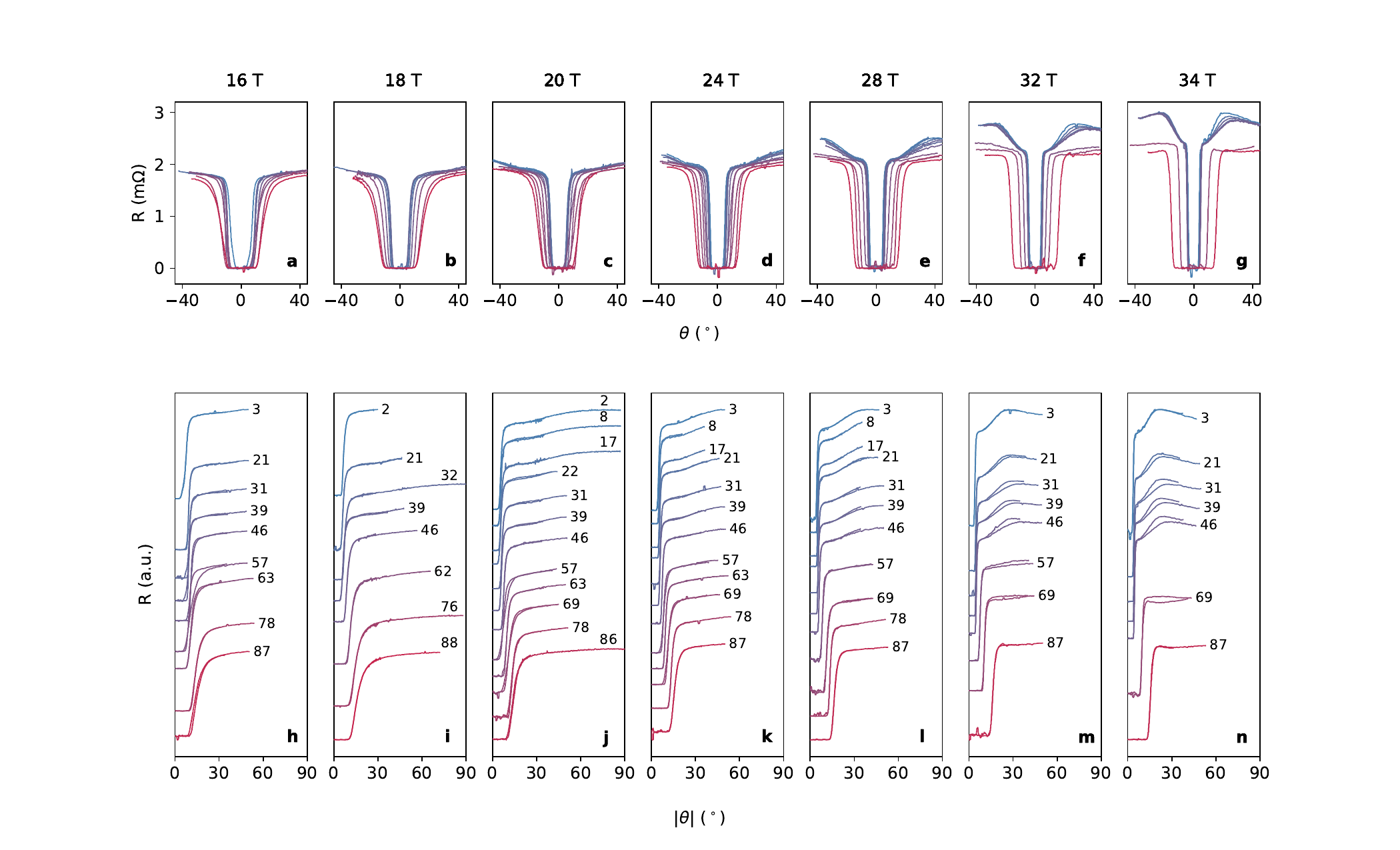}
  	\caption{The resistance is shown as a function of $\theta$ for many values of $\phi$ at various fixed fields,  16 T - 34 T (a-g).  Plots (h-n) show the same data as (a-g), but with each dataset offset by a small amount proportional to $\phi$ in order to highlight the change in feature as a function of $\phi$; each curve is labeled by the value of $\phi$ at which it was taken, in degrees. The data in plots (h-n) has been mirrored about the $b$ axis, i.e. it is shown as a function of $|{\theta}|$. }
  	\label{fig:Rfeature}
	\end{figure*}

In addition to the data shown in Fig. \ref{fig:normalstatefeature}, we present $\theta$-sweeps at various field strengths and various azimuthal angles in Fig. \ref{fig:Rfeature}, so that the evolution of the peak in resistance in the normal state can be seen clearly. For data at higher fields a slight deviation from perfect symmetry can be seen in the resistance, likely due to a minor offset between the axis of rotation and the $b$ axis of the sample. 

\begin{figure}[h]
    	\includegraphics[trim=1cm 0.3cm 1cm 1.2cm, clip=true,width=0.5\textwidth]{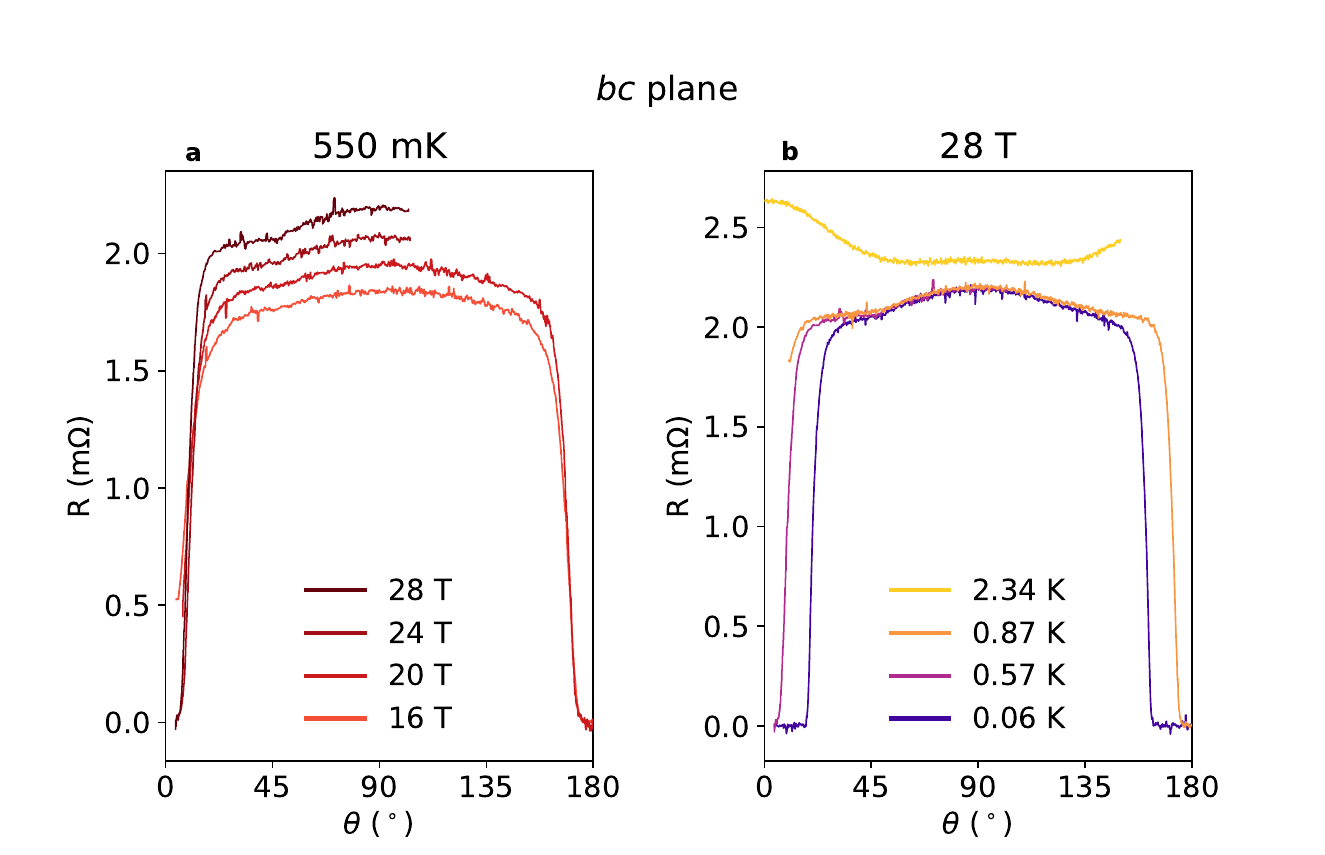}
  	\caption{Resistance as a function of $\theta$ in the $bc$ plane, shown (a) at 550 mK for various field strengths and (b) at 28 T for sweeps taken at various temperatures.  Consistent with the rest of this work, $\theta = 0$ indicates field along the $b$ axis.}
	\label{fig:Rfeaturebc}
	\end{figure}

In Fig. \ref{fig:Rfeaturebc} we focus on data taken with field in the $bc$ plane to show the lack of feature in that plane. From Fig. \ref{fig:Rfeaturebc}(a), we can see that for fields up to 28 T at 550 mK the only resistive maximum occurs with field along $c$; there is no hump at intermediate angles.  In Fig. \ref{fig:Rfeaturebc}(b) we show data taken at 28 T at varying temperatures. While an additional resistive maximum appears for fields along the $b$ axis when the temperature is above $T_c$, there is still no maximum in resistance for field angles between the two crystalline axes.

\begin{figure}
    	\includegraphics[trim=0.8cm 1cm 0.7cm 1.5cm, clip=true,width=0.5\textwidth]{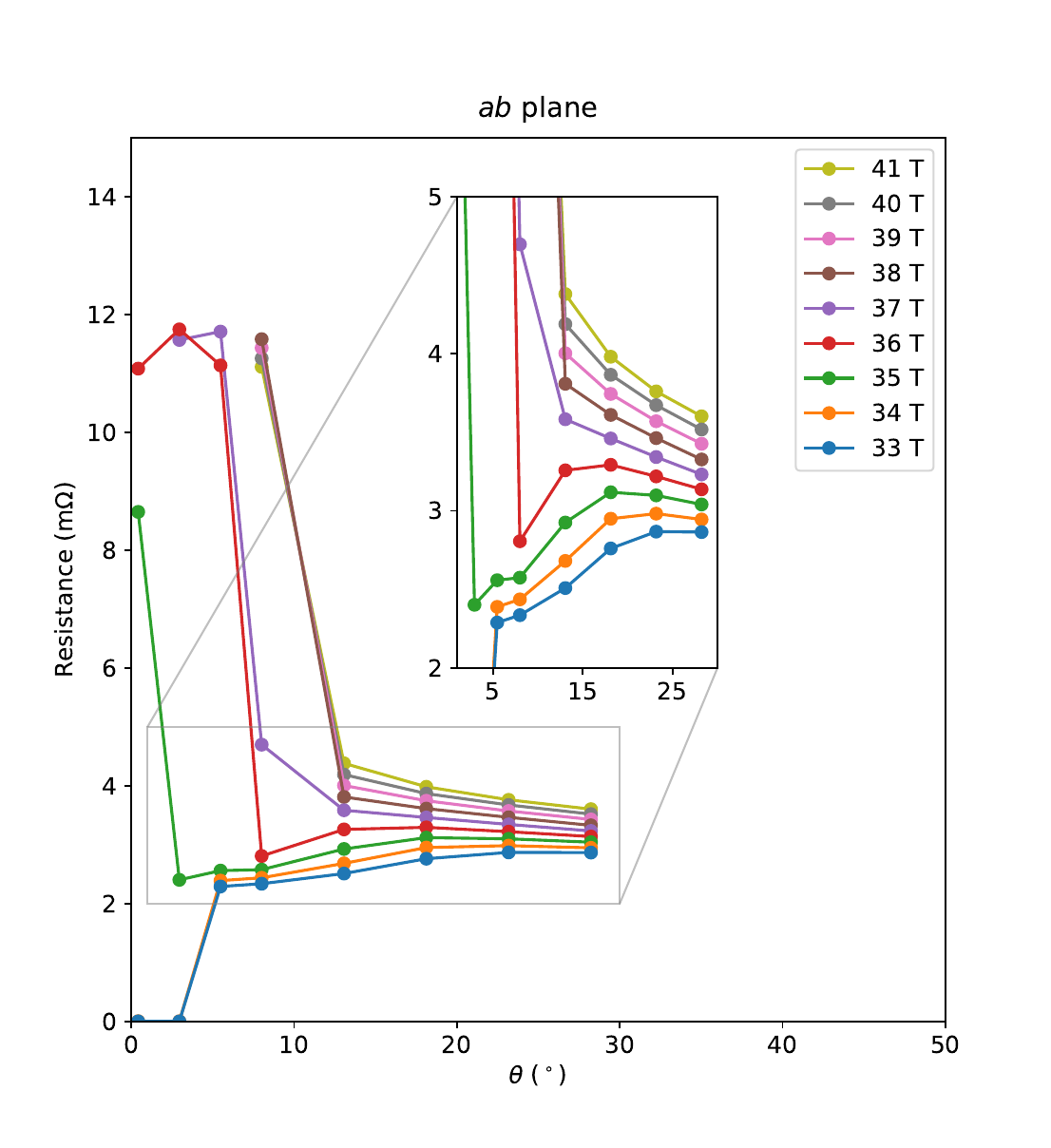}
  	\caption{Data from field sweeps at various angles in the $ab$ plane, replotted to show resistance as a function of angle for fields up to 41 T. The inset shows an enlarged view of the hump in resistance, which is seen up to 36 T.}
  	\label{fig:MMslice}
	\end{figure}

We only measured resistance versus $\theta$ at fields up to 34 T. However, in looking for the boundaries of the FP phase, we also measured resistance as a function of field for several different values of $\theta$ within the $ab$ plane, as shown in Fig. \ref{fig:RvsH}. In Fig. \ref{fig:MMslice}, we replot these data in order to see the behavior of R as a function of $\theta$ for fields up to 41 T.  The feature in resistance can be seen up to 36 T. We fit the plotted 35 T and 36 T data with polynomial functions in order to estimate the angle of maximum resistance for each field, shown as two of the x markers in Fig. \ref{fig:planes}. 

\begin{figure}
    	\includegraphics[trim=0cm 0cm 0cm 0cm, clip=true,width=0.5\textwidth]{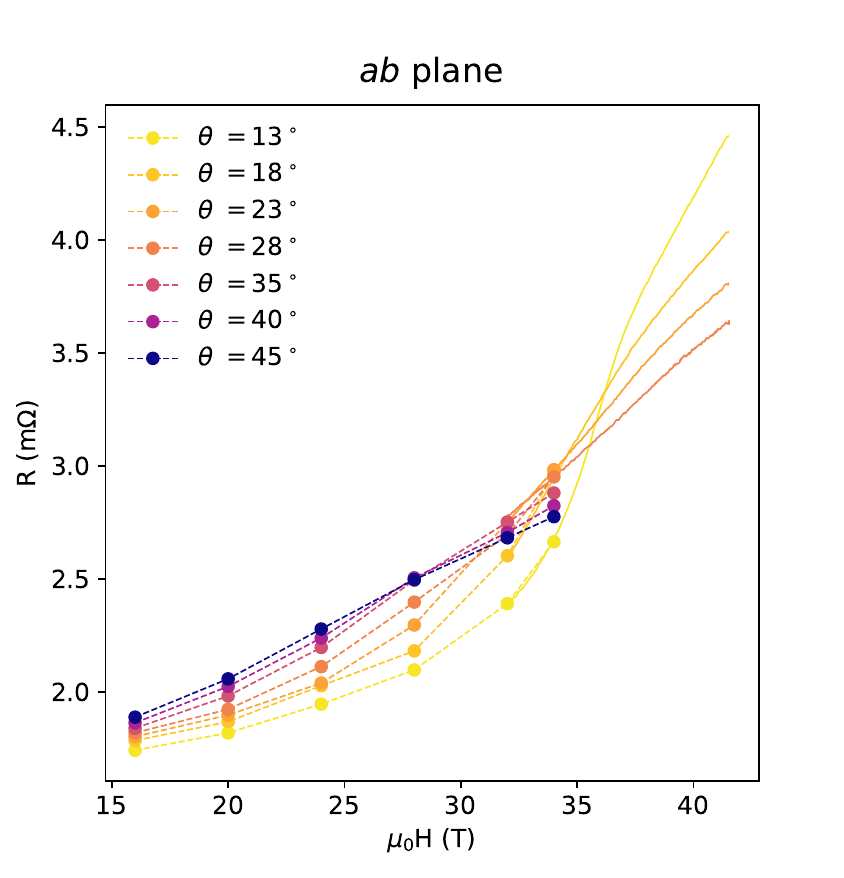}
  	\caption{The resistance of Sample 1 as a function of field strength for various field angles within the $ab$ plane.  The angle $\theta$ indicates the field angle with respect to the crystallographic $b$ axis. Solid lines are field sweeps from Fig. \ref{fig:RvsH}(a). Circles with dashed lines indicate data extracted from angular sweeps at fixed fields.}
  	\label{fig:normalstatefeature2}
	\end{figure}

The peak in resistance is most obvious in the $ab$ plane and seems to disappear as $\phi$ is increased beyond about 50 degrees.  Most of our data were taken using $\theta$-sweeps, but we can replot them to see what would be observed for field sweeps within the $ab$ plane for various values of $\theta$.  Fig. \ref{fig:normalstatefeature2} shows this replotted data combined with some of the field sweep data from Fig. \ref{fig:RvsH}(a). This allows for more direct comparison with the data from UCoGe in Ref. \cite{Bay2014}. We can see that for fields closer to the $b$ axis, the resistance rises much more dramatically as a function of field. Note also that the field-sweep data plotted here are those from Fig. \ref{fig:RvsH}(a) that do \textit{not} undergo a metamagnetic transition in the measured field range; this  increase in resistance is distinct from the huge jump seen at the metamagnetic transition.

\section{Additional normal-state data: hysteresis at the metamagnetic transition}
\label{app:hysteresis}
\begin{figure*}
    	\includegraphics[trim=1cm 1cm 1.5cm 1.5cm, clip=true,width=\textwidth]{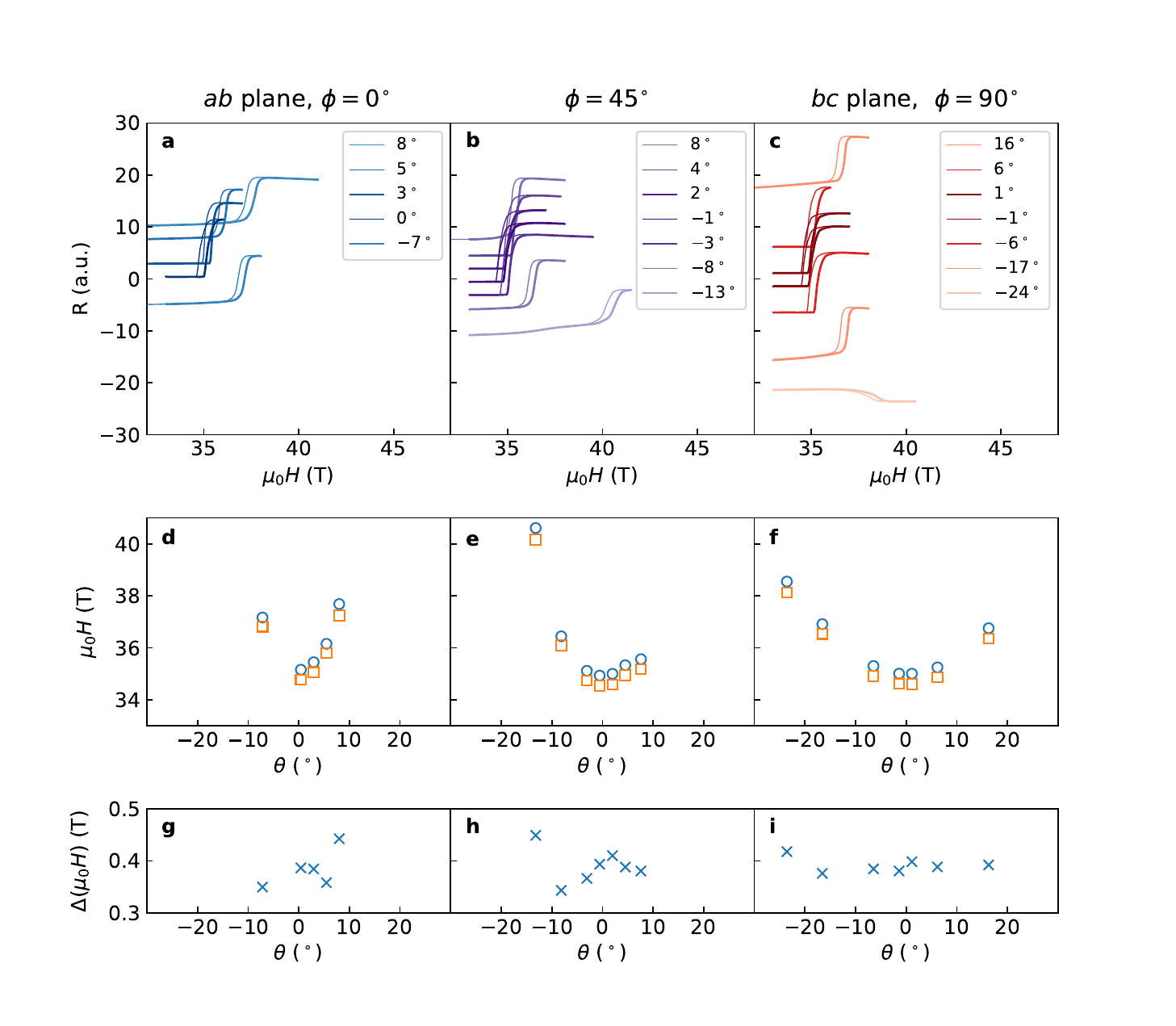}
  	\caption{Hysteresis of the metamagnetic transition as a function of field angle, as measured in resistance.  Each column represents data taken at a fixed $\phi$ with various $\theta$. (a-c) Resistance as a function of field strength; thick lines show field upsweeps and thin lines show field downsweeps.  Legend shows the value of $\theta$ for each field sweep. (d-f) The field at which the metamagnetic transition occurs for various field angles; circles indicate the transition field for upsweeps while squares indicate the transition field for downsweeps. (g-i) The width of the hystereis loop, found by taking the difference of the field of transition between upsweeps and downsweeps.}
  	\label{fig:hysteresis}
	\end{figure*}

With field applied along the $b$ axis, the metamagnetic transition of UTe$_2$ is first-order at temperatures below a critical endpoint; the temperature of the critical endpoint has been reported as 7 K  from magnetoresistance measurements and 11 K from magnetization measurements \cite{Knafo2019, Miyake2019}.  In Fig. \ref{fig:hysteresis} we show how the hysteresis of this transition evolves as a function of field angle, based on our measurements of resistance at 400 mK.  We define the metamagnetic transition field as the field at which the derivative of resistance is greatest.  The width of the hysteresis loop remains relatively steady at $\approx$ 0.4 T for the angles at which we measured.  Interestingly, this includes the measurement at $\theta = -24^{\circ}$ in the $bc$ plane, for which the sample enters the field-polarized superconducting state.  It has previously been observed that the metamagnetic transition appears to form a lower bound of this superconducting state \cite{Ran2019a}.  Rather than indicating that the superconducting transition itself is first-order, this behavior merely underscores the fact that this superconducting state appears only in the field-polarized phase above the metamagnetic transition.

\section{Temperatures within the dilution refrigerator}
\label{app:dilfridge}

Within the dilution refrigerator, we took data at three different approximate temperatures. There was some variation in the initial temperature of each measurement due to temperature instability, which was more pronounced at higher temperatures; in addition, slight heating was observed while performing angular sweeps, presumably due to the motion of the rotator.

For the data denoted as 50 mK, the lowest measured temperature during data taking was 31 mK and the highest was 84 mK.

For the data denoted as 550 mK, the lowest measured temperature during data taking was 529 mK and the highest was 631 mK.

For the data denoted as 900 mK,  the lowest measured temperature during data taking was 811 mK and the highest was 1.09 K.


\begin{thebibliography}{10}
\bibitem{myfootnote}
Note that for Sample 2, the superconducting boundaries actually appear extended at 550 mK compared to 400 mK; it is likely that the nominal 400 mK data for Sample 2 were taken at a slightly elevated temperature due to sample heating, as the sample contacts of Sample 2 degraded somewhat and became higher-resistance before the 400 mK measurements.
\end{thebibliography}
\end{document}